\documentclass[11pt,a4paper]{article}

\usepackage[a4paper,margin=1in]{geometry}
\usepackage[T1]{fontenc}
\usepackage[utf8]{inputenc}
\usepackage{lmodern}
\usepackage{microtype}
\usepackage{graphicx}
\usepackage{multirow}
\usepackage{amsmath,amssymb,amsfonts,bm,mathtools}
\usepackage{mathrsfs}
\usepackage{xcolor}
\usepackage{textcomp}
\usepackage{booktabs}
\usepackage{array}
\usepackage{float}
\usepackage{subcaption}
\usepackage{framed}
\usepackage{enumitem}
\usepackage{placeins}
\usepackage{listings}
\usepackage{titlesec}
\usepackage{hyperref}
\hypersetup{
  colorlinks=true,
  linkcolor=blue,
  citecolor=blue,
  urlcolor=blue,
  linktoc=all,
  bookmarksopen=true,
  bookmarksnumbered=true,
  bookmarksdepth=2,
  pdfstartview=FitH,
  pdftitle={sp-DBA: a general framework for adaptive transform-domain computation},
  pdfauthor={Jingkun Jiang, Pingchuan Deng, Yang Xia}
}

\setlist[enumerate,1]{label=\arabic*.,leftmargin=*,itemsep=0.35em,topsep=0.35em}
\setlist[itemize]{leftmargin=*,itemsep=0.25em,topsep=0.25em}
\titleformat{\section}{\Large\bfseries}{\thesection}{0.65em}{}
\titleformat{\subsection}{\large\bfseries}{\thesubsection}{0.55em}{}
\titlespacing*{\section}{0pt}{2.4ex plus 0.5ex minus 0.2ex}{0.8ex}
\titlespacing*{\subsection}{0pt}{1.8ex plus 0.4ex minus 0.2ex}{0.55ex}
\title{sp-DBA: a general framework for adaptive transform-domain computation}
\author{Jingkun Jiang \quad Pingchuan Deng \quad Yang Xia\textsuperscript{*}\\[6pt]
\small College of Materials Science and Engineering, Hunan University\\
\small Changsha 410082, Hunan, People's Republic of China\\[3pt]
\small Email: \href{mailto:JJJK0506@hnu.edu.cn}{JJJK0506@hnu.edu.cn}
\quad
\textsuperscript{*}Corresponding author: \href{mailto:yxia@hnu.edu.cn}{yxia@hnu.edu.cn}}
\date{}

\begin{document}
\maketitle

\begin{abstract}
Transform-domain methods simplify analysis and computation, making them central to scientific computing and signal processing.
However, existing adaptive strategies often introduce new data structures or require substantial workflow redesign, limiting efficient execution on massively parallel hardware.
Here we present spectral dynamic block activation (sp-DBA), an adaptive acceleration framework for transform-domain workflows.
sp-DBA dynamically activates transform-domain computation blocks as a calculation evolves, concentrating computation where it is needed while retaining numerical accuracy and optimized transform operations.
Across representative workflows in materials science, biology, and optical communications, the demonstrated implementations achieve speedups of up to 28.1-fold for transform-domain updates and up to 8.4-fold overall acceleration; strong- and weak-scaling tests on up to eight GPUs show that the local adaptive update remains effective within distributed FFT workflows.
By introducing execution-level adaptivity into mature transform-domain workflows without rebuilding existing solvers, sp-DBA extends the scale, duration, and complexity of scientific simulations and signal-processing calculations on modern parallel systems.
\end{abstract}

% Put the global contents on its own page for easier navigation in the long preprint.
\clearpage
\setcounter{tocdepth}{2}
\begingroup
\normalsize
\setlength{\parskip}{0pt}
\tableofcontents
\endgroup
\clearpage

\section{Introduction}\label{sec1}

Transform-domain methods are central to scientific computing, engineering, and signal processing because they express fields and signals in representations better suited to analysis and numerical computation~\cite{2000_boyd,2000trefethen}.
Among these methods, Fourier-based workflows are especially widespread.
In a Fourier representation, numerical operations such as differentiation and convolution take simpler forms~\cite{2006_canuto}, while fast Fourier transform (FFT) algorithms substantially reduce the cost of repeated transforms~\cite{1965_fft_ck}.
With advances in parallel computing hardware, highly optimized FFT libraries have been developed for CPUs~\cite{1998_fftw} and GPUs~\cite{cufft}, while distributed FFT implementations have enabled large-scale parallel transforms~\cite{2012_P3Dfft,2020_heffte,cufftmp}.
As a result, Fourier-based workflows are now widely used in areas including materials simulation~\cite{2021_mat_sim2_fftZS,2022_mat_sim1_feFFT}, fluid simulation~\cite{sp_in_cfd}, computational physics~\cite{1996_dft_fftApp,1982_sp_Schrodinger}, medical imaging~\cite{2019_mri_zs}, optical signal processing~\cite{2008_DBP1,2009_DBP2}, and operator learning~\cite{2021_fno,2024_fno_app}.

In addition to advances in transform algorithms and implementations, the increasing computational demands of large-scale and complex scientific problems call for adaptive methods that respond to the evolving structure of fields and signals while avoiding unnecessary computation.
In transform-domain computation, much of the prior work realizes this adaptivity by changing the numerical representation of the field or signal.
Examples include sparse Fourier methods~\cite{2012_sft,2014_sft_review}, wavelet and multiresolution methods~\cite{1999_adp_wavelet,2009_wavelet,2023_adp_wavelet_layerFFT}, and adaptive spectral or reduced-order approaches~\cite{2023_adp_sp2,2022_wavelet_pod}. 
However, in mature Fourier-based workflows built around regular arrays and highly optimized FFT libraries, such changes often require reorganizing the solution data or introducing specialized data structures for adaptivity, limiting direct reuse of existing implementations~\cite{1998_fftw,2012_P3Dfft,2015_pawcm}.
Moreover, this difficulty is amplified on GPUs and other parallel hardware, where irregular execution and data access often reduce the efficiency of adaptive workloads~\cite{2020_gpu_branch_div,2023_irr_coalescing,2021_cufinufft}.

For efficient execution on massively parallel architectures, adaptive methods commonly organize computation into patches or blocks that serve as regular units of parallel work~\cite{2019_AMRex_blockAMR,2017_GAMER2,2018_AMR_gpu2}.
A recent practical realization of this strategy is dynamic block activation (DBA), which organizes computation into blocks and dynamically activates them according to evolving features of the physical model~\cite{2025_dba}.
By preserving a regular blockwise execution pattern on GPUs, DBA offers a straightforward approach to accelerating existing continuum models through adaptivity.
However, applying blockwise activation directly to transform-domain workflows presents two fundamental difficulties.
First, before transformation, the data are represented by local field values, whereas after transformation, they are represented by modes whose contributions span the physical domain, so a transformed-field block does not correspond directly to a local region in the original field~\cite{2000_boyd,2000trefethen,2006_canuto}.
Second, a typical transform-domain workflow tightly couples forward and inverse transforms, the transformed-field update, and other numerical operations, so adaptive decisions must remain consistent with the full workflow to preserve numerical stability and accuracy~\cite{2000trefethen,2006_canuto}.
Consequently, the challenge is to introduce adaptivity without sacrificing the numerical structure and efficient parallel implementations that make mature transform-domain workflows practical.

In this work, we present spectral dynamic block activation (sp-DBA), an adaptive acceleration framework for transform-domain workflows.
sp-DBA applies blockwise activation to transformed-field computation, coordinating adaptive execution with the surrounding transform workflow on parallel hardware.
With only minor changes, we demonstrate the same framework in three representative applications, including polycrystalline coarsening, multicomponent reaction--cross-diffusion, and digital backpropagation, spanning materials science, biology, and optical communications.
Across the three implementations, numerical accuracy is maintained while transform-domain update speedups exceed twentyfold in the strongest configuration.
We further evaluate stage-level cost decomposition and distributed PFC execution on 1--8 GPUs, including strong and weak scaling, to identify where local acceleration persists and where global FFT communication limits end-to-end gains (Fig.~\ref{fig5}; Supplementary Sections~S7 and~S8).
More importantly, this acceleration expands the scientific questions accessible within a fixed computing budget.
Larger domains and longer evolutions provide stronger statistics and clearer separation of intrinsic behavior from finite-size and transient effects, while the same computational savings make richer coupled models and more demanding signal-processing studies accessible.

\section{Methods}\label{sec3}

\begin{figure}[H]
\centering

\includegraphics[width=1.0\linewidth]{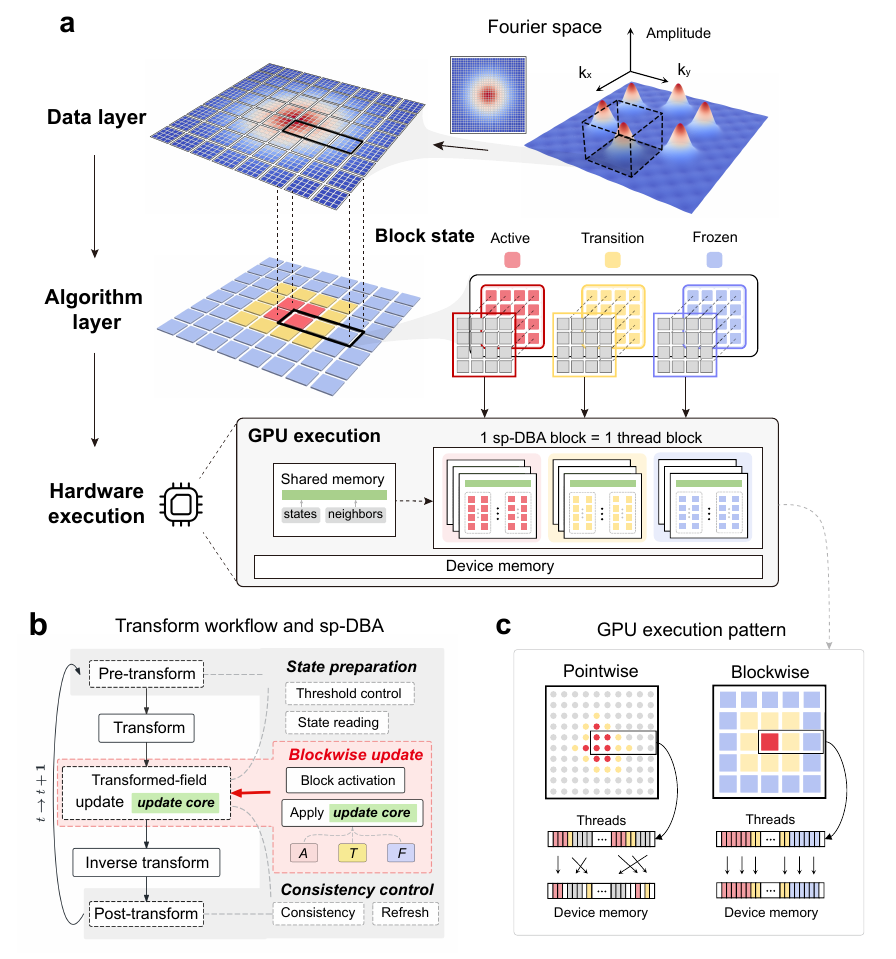}

\caption{
\textbf{Overview of sp-DBA in transform-domain solver workflows.}
\textbf{a,} Data, algorithm, and hardware mapping. The transformed field is partitioned into active, transition, and frozen blocks for full updates, smoothed updates, and update skipping, respectively. Each sp-DBA block maps to one CUDA thread block, with block-state information stored in shared memory.
\textbf{b,} Transform workflow and sp-DBA insertion. sp-DBA contains three modules: state preparation, blockwise update, and consistency control, which respectively manage block states, apply updates adaptively, and maintain numerical consistency.
\textbf{c,} GPU execution patterns. Pointwise activation assigns different update paths to threads within one CUDA thread block, whereas blockwise activation assigns one path to the full thread block.
}
\label{fig1}
\end{figure}

\noindent \textbf{Spectral dynamic block activation.}
sp-DBA is an adaptive acceleration framework for transform-domain workflows.
In the implementations presented here, GPUs serve as a representative parallel platform; their block-based execution model provides a natural mapping for sp-DBA.
Throughout this work, Regular denotes the full-update mode with sp-DBA disabled, in which the model-specific update core is applied to the full transformed field.
With sp-DBA enabled, the same update core is applied according to block state, while the existing transform workflow remains in place.
The framework consists of three modules: state preparation, blockwise update, and consistency control.

For blockwise execution, the transformed-field data are partitioned into sp-DBA blocks, each serving as an activation unit and mapping one-to-one to a CUDA thread block (Fig.~\ref{fig1}a).
State preparation evaluates an application-specific activation measure for each block and compares it with a threshold.
The threshold may be fixed, as in DBA~\cite{2025_dba}, or adjusted dynamically to maintain a target average active-block ratio, which includes both active and transition blocks.
Blocks meeting the threshold condition are marked active.
Among the remaining blocks, those neighboring an active block enter the transition state, whereas the rest are frozen.
The block state then determines the update: active blocks execute the model-specific update core, transition blocks apply a boundary-aware weighted update, and frozen blocks skip the update and retain their stored transformed-field values until reactivation or periodic refresh (Supplementary Section~S1.3).
During GPU execution, the current and neighboring block states, together with the smoothing table, are loaded into shared memory to evaluate the transition weight locally (Supplementary Section~S1.1).

A representative step in a transform-domain workflow includes a pre-transform stage, a forward transform, a transformed-field update, an inverse transform, and a post-transform stage (Fig.~\ref{fig1}b).
For example, in a pseudo-spectral solver, the pre-transform stage evaluates nonlinear or reaction terms in real space, and the forward transform maps them to Fourier space.
At the transformed-field update stage, sp-DBA performs state preparation, blockwise update, and consistency control.
The inverse transform and post-transform stages then proceed as in Regular.
At the hardware level, blockwise activation assigns one update decision to each CUDA thread block, whereas pointwise activation assigns decisions to individual transformed-field values (Fig.~\ref{fig1}c).

These design choices provide four practical benefits: hardware-aligned execution, complementary acceleration, solver compatibility, and simple integration.

\begin{enumerate}
\item \textbf{Acceleration with hardware-aligned block execution.}
Assigning one update decision to each CUDA thread block lets frozen blocks avoid unnecessary updates while keeping all threads in the block on a common execution path (Fig.~\ref{fig1}c).

\item \textbf{Complementary acceleration with optimized transforms.}
Optimized transform implementations, including FFTW~\cite{1998_fftw}, cuFFT~\cite{cufft}, heFFTe~\cite{2020_heffte}, and cuFFTMp~\cite{cufftmp}, reduce the cost of the forward and inverse transforms.
sp-DBA complements these optimized transforms and reduces the number of blocks executing the model-specific update, allowing both forms of acceleration to be combined within the same workflow.

\item \textbf{Compatibility with different numerical solvers.}
sp-DBA operates on each solver's existing transformed-field update while leaving its governing equations, numerical discretization, transformed representation, and transform workflow unchanged.
The same framework is demonstrated in PFC, RCD, and DBP despite differences in their equations and transformed-field layouts (Supplementary Section~S2.1).
By contrast, representative adaptive spectral and wavelet methods change the basis or computational grid and therefore require state transfer or dedicated hierarchical data structures~\cite{2023_adp_sp2,2015_pawcm} (Supplementary Section~S2.3).

\item \textbf{Simple integration into existing solvers.}
Common block operations are implemented once in the shared framework, while each solver reuses its existing update core and supplies the required model-specific data and control settings~\cite{2020_moose}.
Direct integration into the three solvers adds code amounting to about 8--16\% of the corresponding Regular solver core.
Including model-adapter code raises the total to about 13--28\%; the shared framework is excluded from both per-solver counts (Supplementary Section~S2.2).
\end{enumerate}

\section{Results}\label{sec2}

\subsection{Polycrystalline coarsening}\label{subsec1}

\begin{figure}[H]
\centering

\includegraphics[width=1.0\linewidth]{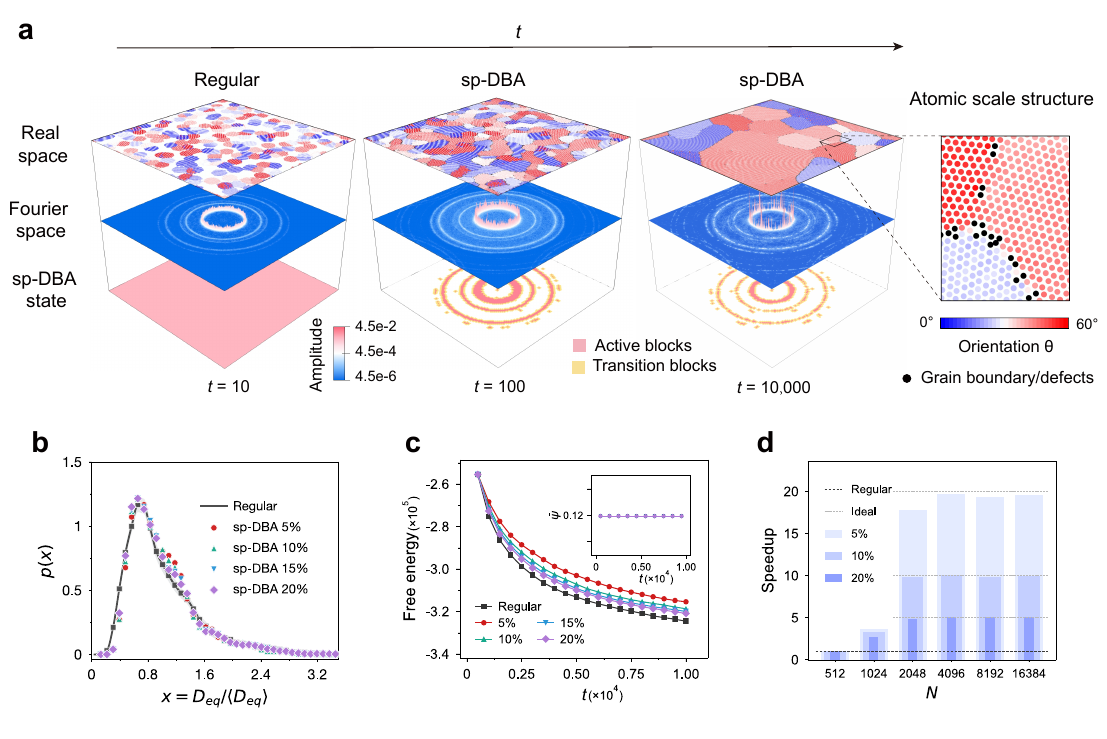}

\caption{
\textbf{Polycrystalline coarsening with sp-DBA.}
\textbf{a,} Layered evolution of XPFC polycrystalline coarsening. From left to right, the columns show snapshots at dimensionless times \(t=10\), \(100\), and \(10,000\), with Regular used for the first and sp-DBA for the latter two, all on a \(2048^2\) grid. From top to bottom, the rows show real-space morphology colored by local crystal orientation \(\theta\), a three-dimensional surface plot of the Fourier-space amplitude with the zero-frequency component omitted, and sp-DBA block states. Red and yellow mark active and transition blocks, respectively, and the inset highlights grain boundaries and atomic-scale defects.
\textbf{b,} Normalized grain-size distributions for Regular and sp-DBA during late-stage coarsening on an \(8192^2\) grid.
\textbf{c,} Free energy \(F[\psi]\) and mean density \(\bar{\psi}\) for Regular and sp-DBA during coarsening on an \(8192^2\) grid.
\textbf{d,} Fourier-space update speedup relative to Regular as a function of grid size. Percentages indicate target average active-block ratios maintained by dynamic threshold control, and dashed lines show the corresponding ideal speedups.
}
\label{fig2}
\end{figure}

We first test sp-DBA in XPFC polycrystalline coarsening, a representative transform-domain workflow linking a lattice-periodic density field to long-time grain-boundary migration. Phase-field-crystal (PFC) models describe crystalline microstructure through a continuous density field \(\psi(\bm{r})\) with periodic order at atomic length scales, capturing lattice ordering, elasticity, defects, and grain boundaries while accessing diffusive time scales beyond direct atomistic simulations~\cite{2002elder,2004elder}. XPFC extends this framework with a two-point correlation kernel \(C_2\) that selects the crystalline length scale and lattice symmetry~\cite{2010_XPFC_prl,2011_XPFC_pre}. As PFC methods have developed, XPFC and related formulations have been applied to crystalline coarsening, structural transformations, and defect evolution~\cite{2012_pfc_review,2023_tmd_xia}. For the XPFC calculations considered here, we use the free energy
\begin{equation}
F[\psi] =\int d\bm{r}\left[\frac12\psi^2-\frac{\omega}{6}\psi^3+\frac{u}{12}\psi^4\right]-\frac12 \iint d\bm{r}\,d\bm{r}'\,\psi(\bm{r}) C_2(|\bm{r}-\bm{r}'|)\psi(\bm{r}').
\end{equation}

In the pseudo-spectral XPFC solver, the local polynomial terms are evaluated from \(\psi(\bm{r})\) in real space, whereas the correlation term is evaluated in Fourier space as \(\hat{C}_2(k)\hat{\psi}(\bm{k})\), where \(\hat{C}_2(k)\) is the Fourier representation of \(C_2\)~\cite{2024_openPFC}. During coarsening, density peaks, defects, and grain boundaries are distributed throughout the physical domain, whereas the dominant Fourier amplitudes concentrate near the Bragg rings associated with the crystalline length scale (Fig.~\ref{fig2}a). Accordingly, sp-DBA reduces or skips updates in blocks away from these rings. This contrast between distributed real-space morphology and localized Fourier amplitudes makes XPFC particularly suitable for testing sp-DBA.

Polycrystalline states are initialized from randomly oriented crystalline seeds. All Fourier blocks are kept active during the rapidly evolving early growth stage; once the polycrystalline structure forms and grain-boundary migration becomes dominant, sp-DBA is enabled for the subsequent stage shown in Fig.~\ref{fig2}a. For each Fourier block \(B_j\), the block energy
\(E_{B_j}=\sum_{\bm{k}\in B_j}|\hat{\psi}(\bm{k})|^2\)
identifies the blocks requiring updates near the Bragg rings. Their gradual evolution allows dynamic threshold control to maintain target average active-block ratios of \(5\%\), \(10\%\), \(15\%\), and \(20\%\) for controlled performance comparisons (Supplementary Section~S3).

To verify agreement with Regular, we compare grain statistics and thermodynamic quantities. Across the tested target ratios, the normalized grain-size distribution (GSD)~\cite{2013_gsd1} closely follows the Regular result (Fig.~\ref{fig2}b), consistent with grain-growth analyses based on equivalent grain diameter~\cite{2014_gsd2,2018_grain_exact}. In addition, the free energy \(F[\psi]\) decreases similarly in the Regular and sp-DBA calculations, while the mean density \(\bar{\psi}\) is conserved (Fig.~\ref{fig2}c). Agreement across grain statistics, free-energy evolution, and mean density confirms that sp-DBA preserves the coarsening statistics and thermodynamic behavior under the tested activation settings.

For performance, lower target active-block ratios produce larger Fourier-space update speedups. For \(N\geq 2048\), the \(20\%\), \(10\%\), and \(5\%\) targets yield approximately \(5\times\), \(10\times\), and \(18\)--\(20\times\) update speedups, respectively (Fig.~\ref{fig2}d). On smaller grids, by contrast, the Bragg-ring region occupies a larger fraction of the Fourier blocks, leaving fewer blocks to skip, while block-state operations account for a larger share of the update time. With increasing grid size, the measured speedups approach the corresponding ideal values, indicating low block-state overhead. This growing benefit at large domains is particularly relevant to coarsening studies, where retaining sufficiently large grain populations is essential for robust late-stage statistics.

\subsection{Multicomponent pattern formation}\label{subsec2}

\begin{figure}[H]
\centering

\includegraphics[width=1.0\linewidth]{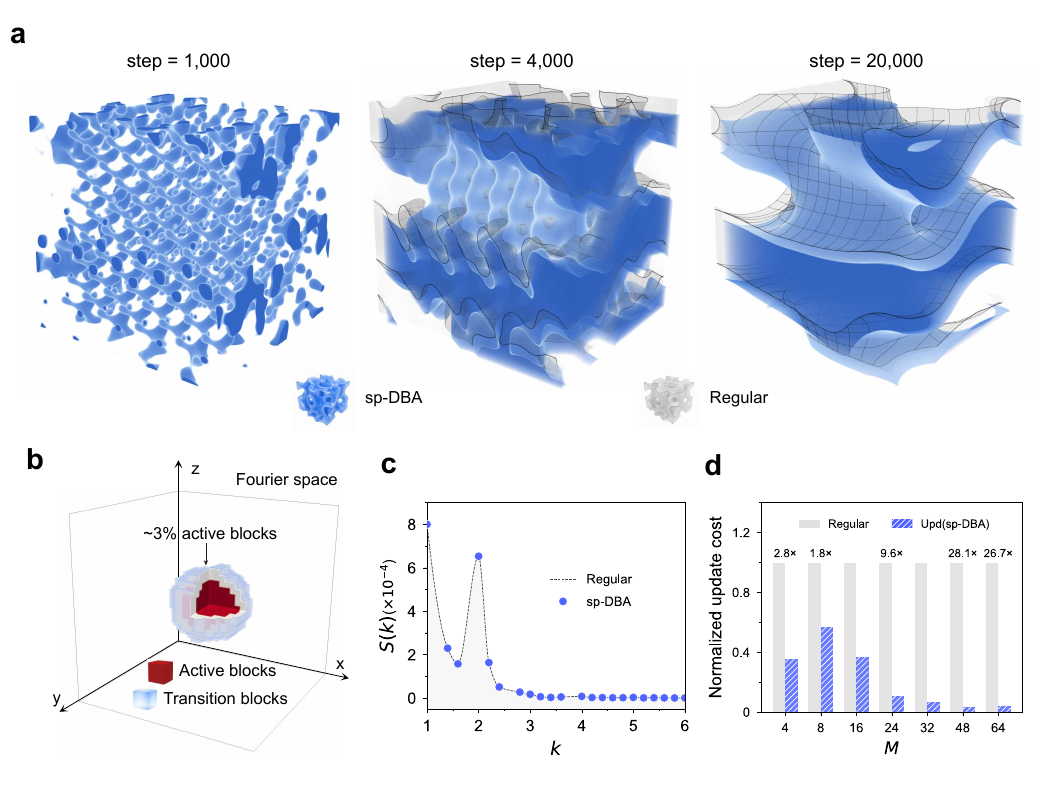}

\caption{
\textbf{Reaction--cross-diffusion pattern formation with sp-DBA.}
\textbf{a,} Real-space \(u_1\) morphology on a \(256^3\) grid with \(M=8\) at 1,000, 4,000, and 20,000 steps. Blue surfaces show sp-DBA; gray wireframes at 4,000 and 20,000 steps show the Regular mode.
\textbf{b,} Fourier-space block states for the same sp-DBA calculation. The average active-block ratio is approximately \(3\%\), counting active and transition blocks. Red and light blue mark active and transition blocks, respectively.
\textbf{c,} Radially averaged structure factor \(S(k)\) of \(u_1\) for Regular and sp-DBA.
\textbf{d,} Normalized Fourier-space update cost as the number of components \(M\) increases, relative to Regular. Labels give the update-stage speedups of sp-DBA with eigendecomposition.
}
\label{fig3}
\end{figure}

We next test sp-DBA in reaction--cross-diffusion (RCD) pattern formation, a multicomponent transform-domain workflow in which each stored Fourier mode carries a coupled update involving \(M\) components. In classical reaction--diffusion systems, reaction kinetics and diffusion can destabilize a homogeneous state and generate spatial patterns from small perturbations~\cite{1952_turing}. Cross diffusion extends this setting by allowing the flux of one component to depend on gradients of other components, providing a mechanism for pattern formation in multicomponent systems~\cite{1979skt,2007_cd_review}. More generally, recent analyses show that non-diagonal diffusion matrices can support diffusion-driven instabilities in reaction--cross-diffusion systems~\cite{2025_RCD}. For the calculations considered here, we use an RCD system in which a constant cross-diffusion matrix couples the \(M\) components,
\begin{equation}
\frac{\partial \bm{u}}{\partial t}=\nabla\cdot(\mathbf{D}\nabla\bm{u})+\mathbf{R}(\bm{u}),
\end{equation}
where \(\bm{u}=(u_1,\ldots,u_M)^T\), \(\mathbf{D}\) is a constant cross-diffusion matrix, and \(\mathbf{R}(\bm{u})\) contains the nonlinear reaction terms (Supplementary Section~S4).

In the Fourier implementation used here, \(\mathbf{R}(\bm{u})\) is evaluated in real space, whereas the cross-diffusion term is updated in Fourier space~\cite{2000_boyd,2000trefethen}. At each Fourier mode \(\bm{k}\), the transformed state is the component vector \(\hat{\bm{u}}(\bm{k})\), and the matrix \(\mathbf{D}\) couples its \(M\) entries. Thus, each stored Fourier mode carries a coupled \(M\times M\) update rather than a scalar multiplication. The eigendecomposition of the constant matrix \(\mathbf{D}\) is computed once before time stepping and reused during the Fourier update~\cite{2020_matrix_rd} (Supplementary Section~S4.3). For the RCD performance comparisons, Regular and sp-DBA use the direct coupled update unless eigendecomposition is stated explicitly.

RCD simulations are initialized with small random perturbations around a homogeneous reference state, after which unstable modes grow and saturate into a labyrinthine pattern. For visualization and spectral comparison, we use component \(u_1\) for the morphology and structure factor \(S(k)\). For block activation, sp-DBA combines all components through the root-mean-square Fourier amplitude \(\mathcal{A}(\bm{k})=(M^{-1}\sum_{i=1}^{M}|\hat{u}_i(\bm{k})|^2)^{1/2}\). Accordingly, the same block state is applied to all \(M\) components; active and transition blocks update the full component vector, whereas frozen blocks skip the coupled operation (Supplementary Section~S4.2).

To assess agreement with Regular, we compare real-space morphology and wavelength selection. As the instability grows and saturates, the \(u_1(\bm{r})\) morphology from sp-DBA closely matches the Regular result, with the blue sp-DBA surfaces overlapping the gray Regular wireframes at 4,000 and 20,000 steps (Fig.~\ref{fig3}a). In Fourier space, active and transition blocks concentrate in the low-wavenumber band where the unstable modes grow, leaving most stored Fourier blocks frozen (Fig.~\ref{fig3}b). This localization is consistent with the finite instability band that selects the dominant wavelength~\cite{2025_RCD,1977swift}. To quantify wavelength selection, we compute the radially averaged structure factor \(S(k)=\langle S(\bm{k})\rangle_{|\bm{k}|=k}\) from \(u_1\). The peak positions and amplitudes agree closely between Regular and sp-DBA (Fig.~\ref{fig3}c). Outside the unstable band, high-wavenumber modes decay rapidly, so the required updates are concentrated at low wavenumbers (Supplementary Section~S4.2). The agreement in morphology and \(S(k)\) shows that sp-DBA preserves wavelength selection even though most high-wavenumber blocks are frozen.

For performance, increasing \(M\) makes each frozen block skip a larger coupled Fourier-space update. Regular applies the \(M\times M\) operation at every stored wavenumber, whereas sp-DBA executes it only in active and transition blocks. Relative to Regular, sp-DBA with eigendecomposition gives update-stage speedups of \(9.6\times\) at \(M=24\), \(28.1\times\) at \(M=48\), and \(26.7\times\) at \(M=64\) (Fig.~\ref{fig3}d). At smaller \(M\), block-state reading and updates account for a larger fraction of the update-stage cost, resulting in smaller measured speedups. These measurements show that block activation removes a large fraction of the coupled update cost, directly targeting the growing computational burden of high-component models. Fig.~\ref{fig5}d and Supplementary Section~S4.3 separate this contribution from eigendecomposition.

\subsection{Digital backpropagation}\label{subsec3}

\begin{figure}[H]
\centering

\includegraphics[width=1.0\linewidth]{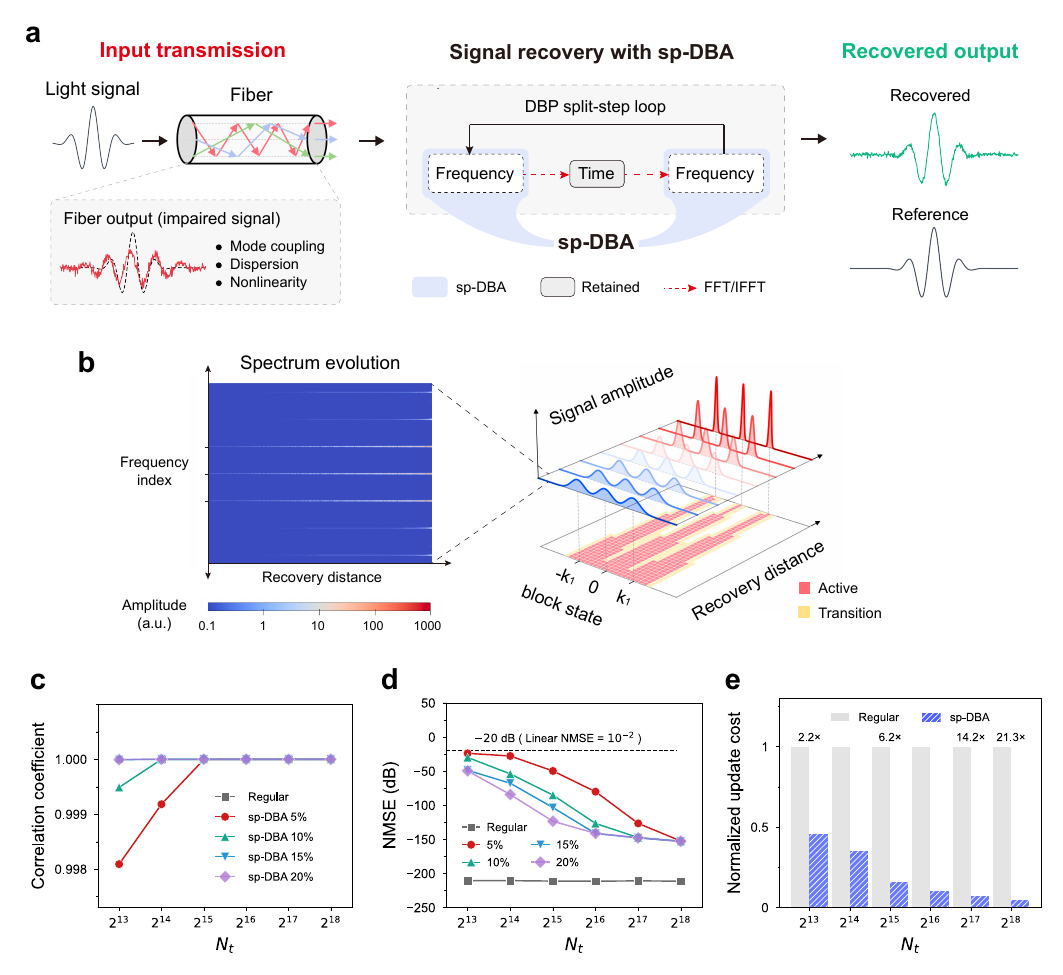}

\caption{
\textbf{Digital backpropagation with sp-DBA.}
\textbf{a,} Split-step digital backpropagation for optical signal recovery, with sp-DBA applied to the frequency-domain linear update.
\textbf{b,} Evolution of the signal spectrum and block states, together with a schematic view of waveform recovery. Red and yellow mark active and transition blocks, respectively.
\textbf{c,} Field correlation coefficient of the recovered signal relative to the transmitted reference versus signal length \(N_t\).
\textbf{d,} Normalized mean-squared error (NMSE) of the recovered signal versus \(N_t\). In \textbf{c,d}, percentages indicate target average active-block ratios, and the dashed line in \textbf{d} marks the \(-20\) dB reference.
\textbf{e,} Normalized frequency-domain update cost for Regular and sp-DBA versus \(N_t\). The default sp-DBA setting targets a \(10\%\) average active-block ratio, and labels give the update-stage speedups.
}
\label{fig4}
\end{figure}

To test sp-DBA in an engineering workflow dominated by repeated frequency-domain updates, we apply it to space-division-multiplexed digital backpropagation (SDM-DBP) for optical signal recovery (Fig.~\ref{fig4}a). In SDM links, signals propagate through multiple spatial modes or fiber cores, increasing transmission capacity while producing coupled complex fields affected by dispersion, mode coupling, attenuation, and Kerr nonlinearity~\cite{2013_sdmdbp,2012_sdm_matrix,2013_manakov}. At the receiver, DBP compensates these impairments by numerically reversing signal propagation through split-step updates that alternate frequency-domain linear propagation and time-domain nonlinear correction. Within this split-step procedure, increasing the number of propagation steps generally improves the approximation of inverse propagation, but each added step repeats both updates; longer signal windows and more spatial channels further increase the computational workload~\cite{2008_DBP1,2009_DBP2,2020_NC_DBP}. Accordingly, the repeated split-step workload has motivated reduced-complexity methods in optical signal processing~\cite{2011_simplified_dbp,2017_dbp_limit_review}.
For the coupled field envelope \(\bm{E}=[E_1,\ldots,E_M]^T\), the SDM-DBP model used here is
\begin{equation}
\frac{\partial \bm{E}}{\partial z}=(\hat{D}+\hat{N})\bm{E}=\left[\left(-\frac{\alpha}{2}\bm{I}-i\frac{\beta_2}{2}\frac{\partial^2}{\partial t^2}\bm{I}+i\bm{K}\right)+i\gamma\frac{8}{9}\left(\sum_{m=1}^{M}|E_m|^2\right)\bm{I}\right]\bm{E},
\end{equation}
where the linear operator \(\hat{D}\) includes attenuation \(\alpha\), chromatic dispersion \(\beta_2\), and mode coupling \(\bm{K}\), while the nonlinear operator \(\hat{N}\) represents the Kerr response with coefficient \(\gamma\) driven by the total modal power (Supplementary Section~S5.1).

In the numerical implementation, sp-DBA is applied to the repeated frequency-domain evaluation of \(\hat{D}\), while \(\hat{N}\) is evaluated in the time domain~\cite{1984_ssfm,1968_hfh_steps,2003_ssfm2}. The input field consists of compact Gaussian wave packets with different carrier-frequency offsets, producing spectral bands that remain localized during backpropagation (Fig.~\ref{fig4}b). For block activation, sp-DBA uses the total modal spectral power \(\mathcal{P}(\omega)=\sum_{m=1}^{M}|\tilde{E}_m(\omega)|^2\). The resulting frequency-block states therefore follow the spectral bands occupied by all modes (Supplementary Section~S5.3).

To assess recovery quality, we compare each recovered field with the transmitted reference using the field correlation coefficient and normalized mean-squared error (NMSE) over the full multimode complex field.
NMSE is commonly used to evaluate recovery in optical DBP and nonlinear-compensation studies~\cite{2008_DBP1,2020_NC_DBP} (Supplementary Section~S5.3). Across the tested signal lengths \(N_t\), the sp-DBA field correlation coefficient remains close to unity, with the lowest plotted value around 0.998 and the larger-\(N_t\) cases approaching the Regular result (Fig.~\ref{fig4}c). Likewise, the sp-DBA NMSE remains at or below the \(-20\) dB reference, corresponding to a linear NMSE no larger than \(10^{-2}\), whereas the Regular calculation gives near-machine-precision recovery for this controlled input (Fig.~\ref{fig4}d). Overall, the correlation and NMSE results show that sp-DBA preserves signal-recovery accuracy under the tested activation settings.

For performance, increasing \(N_t\) expands the sampled frequency grid around these occupied bands, leaving more off-band blocks frozen during the linear update. Consequently, at the default \(10\%\) target average active-block ratio, the frequency-domain update-stage speedup increases from \(2.2\times\) at \(N_t=2^{13}\) to \(21.3\times\) at \(N_t=2^{18}\) (Fig.~\ref{fig4}e). Reduced-complexity DBP methods commonly lower cost by narrowing the processed spectrum through sub-bands or bandwidth restrictions~\cite{2017_dbp_limit_review,2012_subband_dbp}, simplifying the propagation model~\cite{2011_simplified_dbp}, or changing the split-step schedule~\cite{2003_ssfm2,2011_logstep_dbp}. 
As the sampled spectrum expands beyond the occupied bands, sp-DBA becomes increasingly effective for longer signal windows, where repeated frequency-domain updates otherwise impose a growing computational cost.

\subsection{Performance and scalability}

\FloatBarrier

\begin{figure}[H]
\centering

\includegraphics[width=1.0\linewidth]{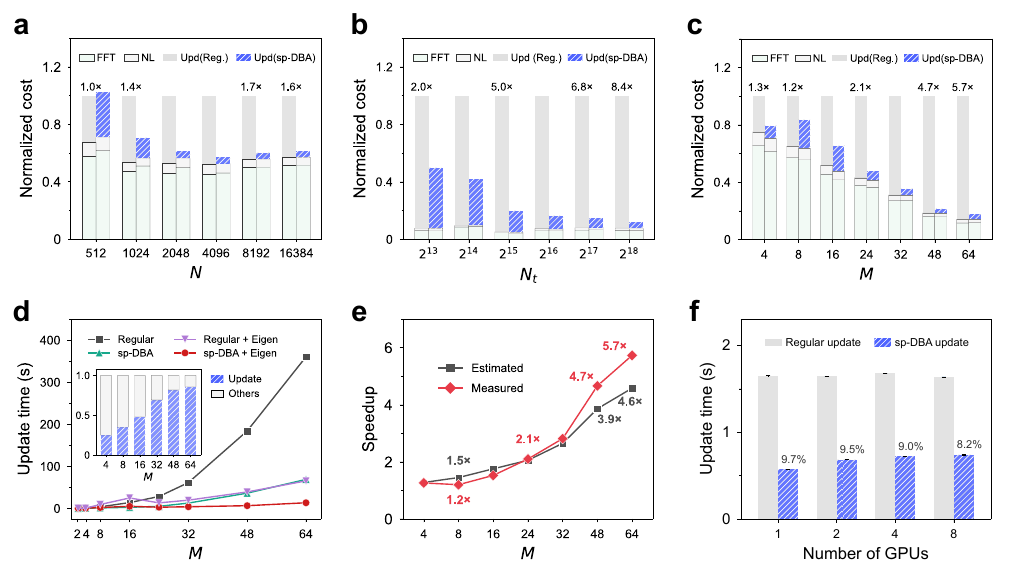}

\caption{
\textbf{Performance and scalability of sp-DBA.}
Unless noted otherwise, dynamic threshold control targets a \(10\%\) average active-block ratio, and costs in \textbf{a--c} are normalized to the corresponding Regular baseline. FFT, NL, and Upd. denote transform operations, model-specific nonlinear calculations, and transformed-field updates.
\textbf{a,} PFC cost decomposition as grid size \(N\) increases.
\textbf{b,} DBP cost decomposition as signal length \(N_t\) increases at \(M=32\).
\textbf{c,} RCD cost decomposition as component number \(M\) increases, comparing Regular and sp-DBA with eigendecomposition.
\textbf{d,} RCD update time for Regular and sp-DBA with and without eigendecomposition. The inset shows the Regular runtime fractions of the update and all other stages.
\textbf{e,} Block-activation estimate and measured combined RCD end-to-end acceleration as a function of \(M\), both relative to Regular.
\textbf{f,} PFC update-stage weak scaling at approximately fixed per-GPU workload. Bars show means over three runs, error bars indicate \(\pm 1\) standard deviation, and labels give measured average active-block ratios.
}
\label{fig5}
\end{figure}

With transform operations and other model steps retained, end-to-end acceleration depends on the share of the Regular runtime spent in the transformed-field update (Fig.~\ref{fig5}a--c). In PFC, the update-stage speedup approaches \(10\times\) at large grids, although the two-dimensional FFT and real-space stages still account for a substantial fraction of the runtime (Fig.~\ref{fig5}a). In DBP, the resulting update reduction allows end-to-end acceleration to reach \(8.4\times\) (Fig.~\ref{fig5}b). In RCD, the coupled update accounts for a larger share of the Regular runtime as \(M\) increases. Accordingly, the end-to-end acceleration of sp-DBA with eigendecomposition relative to Regular shows an overall upward trend (Fig.~\ref{fig5}c).

RCD also demonstrates the complementary roles of sp-DBA and eigendecomposition. Eigendecomposition lowers the cost of the matrix operation at each updated Fourier mode, whereas sp-DBA reduces the number of blocks that perform it (Supplementary Section~S4.3). At \(M=48\), combining the two gives the lowest update time and a \(28.1\times\) update-stage speedup relative to Regular (Figs.~\ref{fig3}d and \ref{fig5}d). For the complete calculation, we compare the measured end-to-end acceleration of the combined configuration with a block-activation estimate (Fig.~\ref{fig5}e). The estimate uses the Regular update-stage cost share and average active-block ratio and excludes the eigendecomposition gain. At small \(M\), block-control overhead is more visible, and the measured acceleration is close to or below the estimate. At larger \(M\), eigendecomposition provides a larger additional reduction, and the measured curve rises above the estimate.

Finally, we evaluate the sp-DBA update stage within a distributed PFC FFT workflow. Under weak scaling, the per-GPU workload remains approximately fixed, the measured average active-block ratio stays near the \(10\%\) target, and the update times of both Regular and sp-DBA remain stable from one to eight GPUs. sp-DBA also reduces the per-GPU update time by about \(2\)--\(3\times\) (Fig.~\ref{fig5}f). Supplementary strong-scaling tests and distributed cost decompositions likewise show lower sp-DBA update times (Supplementary Section~S8). The global transpose and all-to-all communication remain unchanged, but the update-stage reduction persists across the tested distributed configurations.

\section{Discussion}\label{sec4}

sp-DBA differs from established forms of transform-domain adaptivity in what it adapts.
Existing approaches commonly obtain adaptivity by changing the active spectral set~\cite{2012_sft,2014_sft_review}, the expansion order or basis parameters~\cite{2023_adp_sp2, 2021_p_adaptive}, or the computational grid and coefficient hierarchy~\cite{1999_adp_wavelet,2023_adp_wavelet_layerFFT,2015_pawcm}.
sp-DBA instead introduces adaptivity at the level of execution, leaving the transformed representation and transform workflow unchanged while dynamically controlling where computation is performed.
This separation shifts adaptivity from solver-specific numerical redesign to a reusable execution strategy for transform-domain computation.
The three demonstrations show that this execution-level principle carries across substantially different scientific and engineering problems.

It is notable that the acceleration provided by sp-DBA increases both the spatial scale and evolution time accessible within the same computing budget.
For grain-coarsening studies, this broader computational reach is scientifically important because reliable statistics require domains large enough to retain sufficient grains and simulations long enough to distinguish transient dynamics from steady or asymptotic behavior.
Access to larger domains and longer times therefore enables more reliable grain-size distributions, reduces finite-size uncertainty, and strengthens comparisons among theoretical predictions, numerical simulations, and grain-growth behavior observed in real materials~\cite{2017_ultralarge_grain}.

The same acceleration also makes more complex models accessible.
As shown by the multicomponent RCD model, each Fourier update becomes more demanding as the number of interacting components increases.
By reducing repeated transformed-field work, sp-DBA lowers the computational barrier to studying pattern formation in systems with more interacting species~\cite{2025_RCD} and exploring wider ranges of reaction networks, diffusion parameters, and pattern-forming regimes~\cite{2016_turing_networks}.
In optical communications, simulations of multimode signal transmission likewise become more demanding as the number of modes and operating conditions increases~\cite{2020_mmf_multigpu}.
These examples point to richer models and broader parameter regimes as natural uses of the saved computation.

The present results also define the regimes in which sp-DBA is most effective.
The largest gains occur when the transformed-field update accounts for a substantial fraction of the computational cost, spectral activity remains concentrated over time, and the active-block ratio remains low.
End-to-end acceleration is nevertheless limited by the share of Regular runtime spent in the update stage, as transform operations, communication, and other model steps still contribute to runtime, consistent with Amdahl's law~\cite{1967amdahl}.
A second limitation is that the activation criteria must identify the spectral regions whose updates affect the solution.
In the three demonstrations, the dominant changes remain confined to limited spectral regions or distinct bands.
More general optical-fiber propagation regimes, however, may involve nonlinear phase evolution with little amplitude change and four-wave mixing that transfers energy across frequencies~\cite{2024_sdm_nl_numericalAlg,2023_fwm}.

Looking ahead, phase-aware criteria, more conservative activation rules, and refresh safeguards offer natural extensions toward regimes with weak amplitude signatures, while dynamic load balancing is the most immediate extension of the distributed implementation. Other directions include block layouts designed for specialized transform hardware and activation rules suited to additional transform families. More broadly, sp-DBA points toward a hardware-aware form of adaptive scientific computing in which high-accuracy numerical methods are paired with controlled adaptive execution to achieve higher performance and broader computational reach.

\FloatBarrier

\section*{Acknowledgements}
We thank Y. Wu for useful discussions.

\section*{Author contributions}
J.J. and Y.X. conceived the study and designed the sp-DBA framework. J.J. implemented the code, performed the simulations, analyzed the data, and drafted the manuscript. P.D. contributed to validation and manuscript review and editing. Y.X. supervised the study and revised the manuscript. All authors reviewed and approved the manuscript.

\section*{Competing interests}
The authors declare no competing interests.

\phantomsection
\addcontentsline{toc}{section}{References}
\bibliographystyle{unsrt}
\bibliography{Reference}

\clearpage
\begin{center}
{\LARGE\bfseries Supplementary Information}
\end{center}
\vspace{0.9em}
\phantomsection
\addtocontents{toc}{\protect\setcounter{tocdepth}{2}}
\addcontentsline{toc}{section}{Supplementary Information}

\begingroup
\normalsize
\setlength{\parskip}{0.22em}
\newcommand{\suppTOCsec}[3]{\noindent\hyperref[#1]{\textbf{#2\quad #3}}\dotfill\pageref*{#1}\par}
\newcommand{\suppTOCsub}[3]{\noindent\hspace*{1.5em}\hyperref[#1]{#2\quad #3}\dotfill\pageref*{#1}\par}
\suppTOCsec{sec:core_mechanisms}{S1}{sp-DBA framework}
\suppTOCsub{supp:s1.1}{S1.1}{Boundary-aware transition smoothing}
\suppTOCsub{supp:s1.2}{S1.2}{Neighbor activation in half-spectrum FFT layouts}
\suppTOCsub{supp:s1.3}{S1.3}{Dynamic threshold control and periodic refresh}
\suppTOCsec{sec:solver_integration}{S2}{sp-DBA integration with transform-domain solvers}
\suppTOCsub{supp:s2.1}{S2.1}{Integration procedure}
\suppTOCsub{supp:s2.2}{S2.2}{Implementation overhead}
\suppTOCsub{supp:s2.3}{S2.3}{Comparison with adaptive spectral and wavelet methods}
\suppTOCsec{sec:si_pfc}{S3}{XPFC polycrystalline coarsening}
\suppTOCsub{supp:s3.1}{S3.1}{Governing equations}
\suppTOCsub{supp:s3.2}{S3.2}{Numerical scheme and simulation configuration}
\suppTOCsub{supp:s3.3}{S3.3}{Grain reconstruction and system-size comparison}
\suppTOCsec{sec:si_rcd}{S4}{Reaction--cross-diffusion pattern formation}
\suppTOCsub{supp:s4.1}{S4.1}{Governing equations}
\suppTOCsub{supp:s4.2}{S4.2}{Numerical scheme and simulation configuration}
\suppTOCsub{supp:s4.3}{S4.3}{Matrix eigendecomposition for GPU acceleration}
\suppTOCsec{sec:si_dbp}{S5}{Digital backpropagation}
\suppTOCsub{supp:s5.1}{S5.1}{Governing equations}
\suppTOCsub{supp:s5.2}{S5.2}{Split-step Fourier implementation}
\suppTOCsub{supp:s5.3}{S5.3}{Simulation configuration and recovery metrics}
\suppTOCsec{sec:si_validation}{S6}{Numerical validation}
\suppTOCsub{supp:s6.1}{S6.1}{Conservation and free-energy evolution}
\suppTOCsub{supp:s6.2}{S6.2}{Spectral accuracy and numerical error}
\suppTOCsub{supp:s6.3}{S6.3}{Dealiasing compatibility}
\suppTOCsec{sec:si_additional_performance}{S7}{Single-GPU performance evaluation}
\suppTOCsub{supp:s7.1}{S7.1}{PFC cost decomposition and acceleration}
\suppTOCsub{supp:s7.2}{S7.2}{DBP cost decomposition and acceleration}
\suppTOCsec{sec:si_distributed}{S8}{Multi-GPU sp-DBA with distributed FFTs}
\suppTOCsub{supp:s8.1}{S8.1}{Distributed FFT workflow}
\suppTOCsub{supp:s8.2}{S8.2}{Strong and weak scaling}
\suppTOCsub{supp:s8.3}{S8.3}{CUDA-aware MPI communication}
\suppTOCsub{supp:s8.4}{S8.4}{Activation stability under distributed execution}
\suppTOCsec{sec:si_visualization}{S9}{Visualization}
\endgroup
\clearpage

\setcounter{section}{0}
\setcounter{subsection}{0}
\setcounter{equation}{0}
\setcounter{figure}{0}
\setcounter{table}{0}
\renewcommand{\thesection}{S\arabic{section}}
\renewcommand{\thesubsection}{S\arabic{section}.\arabic{subsection}}
\renewcommand{\theequation}{S.\arabic{equation}}
\renewcommand{\thefigure}{S\arabic{figure}}
\renewcommand{\thetable}{S\arabic{table}}

\renewcommand{\theHsection}{supp.\arabic{section}}
\renewcommand{\theHsubsection}{supp.\arabic{section}.\arabic{subsection}}
\renewcommand{\theHequation}{supp.\arabic{equation}}
\renewcommand{\theHfigure}{supp.\arabic{figure}}
\renewcommand{\theHtable}{supp.\arabic{table}}

\section{sp-DBA framework}
\label{sec:core_mechanisms}
% ======================================================================

sp-DBA uses boundary-aware transition smoothing, half-spectrum neighbor activation, dynamic threshold control, and periodic refresh to manage block states and update boundaries in transformed fields. Hereafter, Regular refers to the full-update mode with sp-DBA disabled.

\subsection{Boundary-aware transition smoothing}\label{supp:s1.1}

When the solver update is skipped in frozen blocks, their transformed-field values are retained from the previous step. A moving boundary can therefore appear between updated and skipped regions in the one-step update increment. To reduce this boundary discontinuity, sp-DBA applies a smooth weight to the newly applied increment,
\begin{equation} \hat{q}_{n+1}(\bm{k}) = \hat{q}_{n}(\bm{k}) + \sigma(\bm{k})\Delta \hat{q}_{\mathrm{raw}}(\bm{k}), \end{equation}
where $\hat{q}$ is a generic transformed variable, $\Delta \hat{q}_{\mathrm{raw}}$ is the raw update increment, and $\sigma(\bm{k})$ is the transition weight. The weight is applied only to the increment, so the stored values in frozen blocks are not damped by the smoothing operation~\cite{2000_boyd,2006_canuto}.

For a block with edge length $B$, the one-dimensional taper is defined as
\begin{equation} w(i)=\frac{1}{2} \left[ 1-\cos\!\left(\frac{\pi i}{B-1}\right) \right], \qquad i=0,1,\ldots,B-1 . \end{equation}
The current block state and neighboring block states are loaded into shared memory, and the local transition weight is assembled from the frozen-neighbor pattern. In two dimensions, let $\chi_L,\chi_R,\chi_D,\chi_U\in\{0,1\}$ indicate whether the left, right, lower and upper neighboring blocks are frozen. The local weight is then
\begin{equation} \sigma_{2D}(i,j;t)= \left[w(i)\right]^{\chi_L} \left[w(B-1-i)\right]^{\chi_R} \left[w(j)\right]^{\chi_D} \left[w(B-1-j)\right]^{\chi_U}. \end{equation}
The same construction extends to three dimensions by adding the front and back factors. Only the one-dimensional taper and block-state metadata are stored, while the multidimensional weight is assembled locally during the update.

\begin{figure}[H]
\centering
\includegraphics[width=1.0\linewidth]{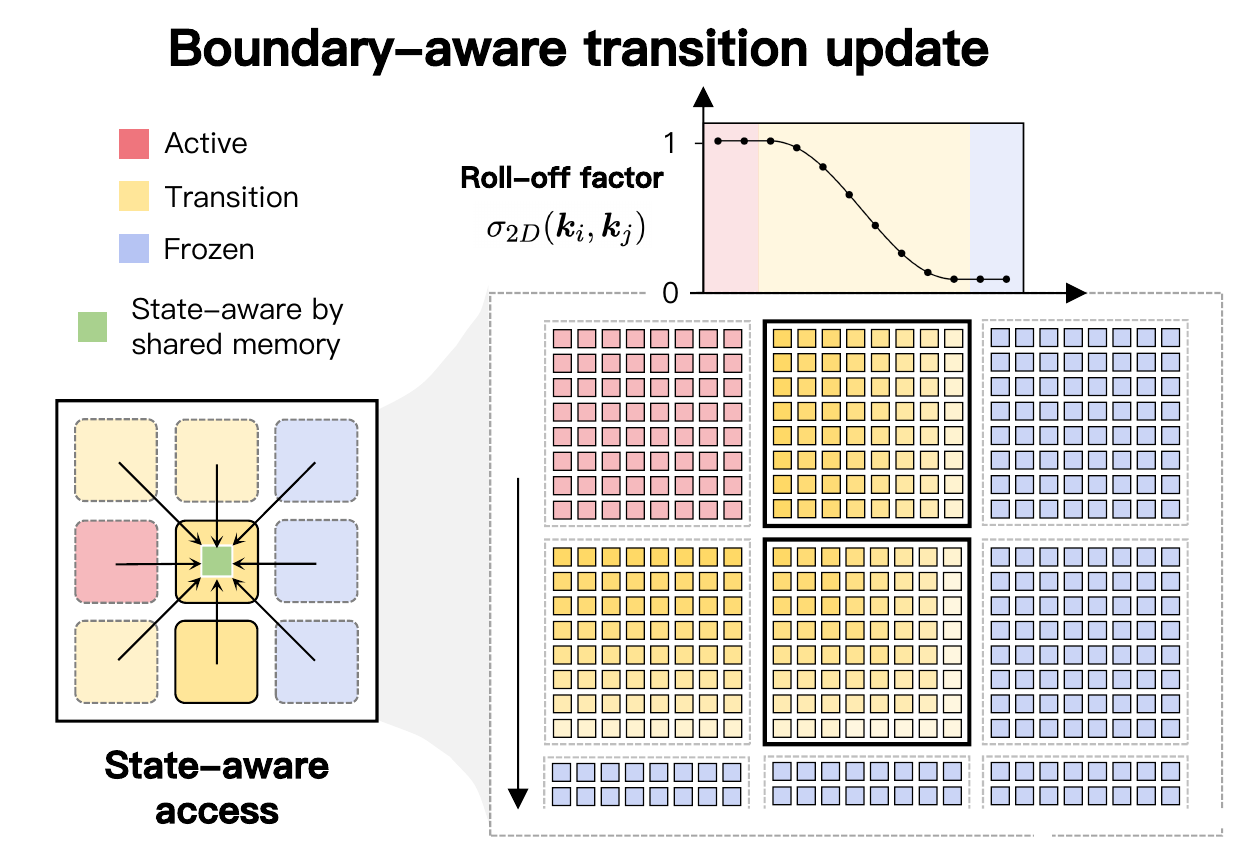}
\caption{Boundary-aware transition smoothing. Neighboring block states are loaded through shared memory to form a local roll-off weight. Active blocks receive the full update increment, transition blocks receive a weighted increment, and frozen blocks retain their previous transformed-field values.}
\label{fig:s1_smoothing_mechanism}
\end{figure}

\begin{figure}[H]
\centering
\includegraphics[width=1.0\linewidth]{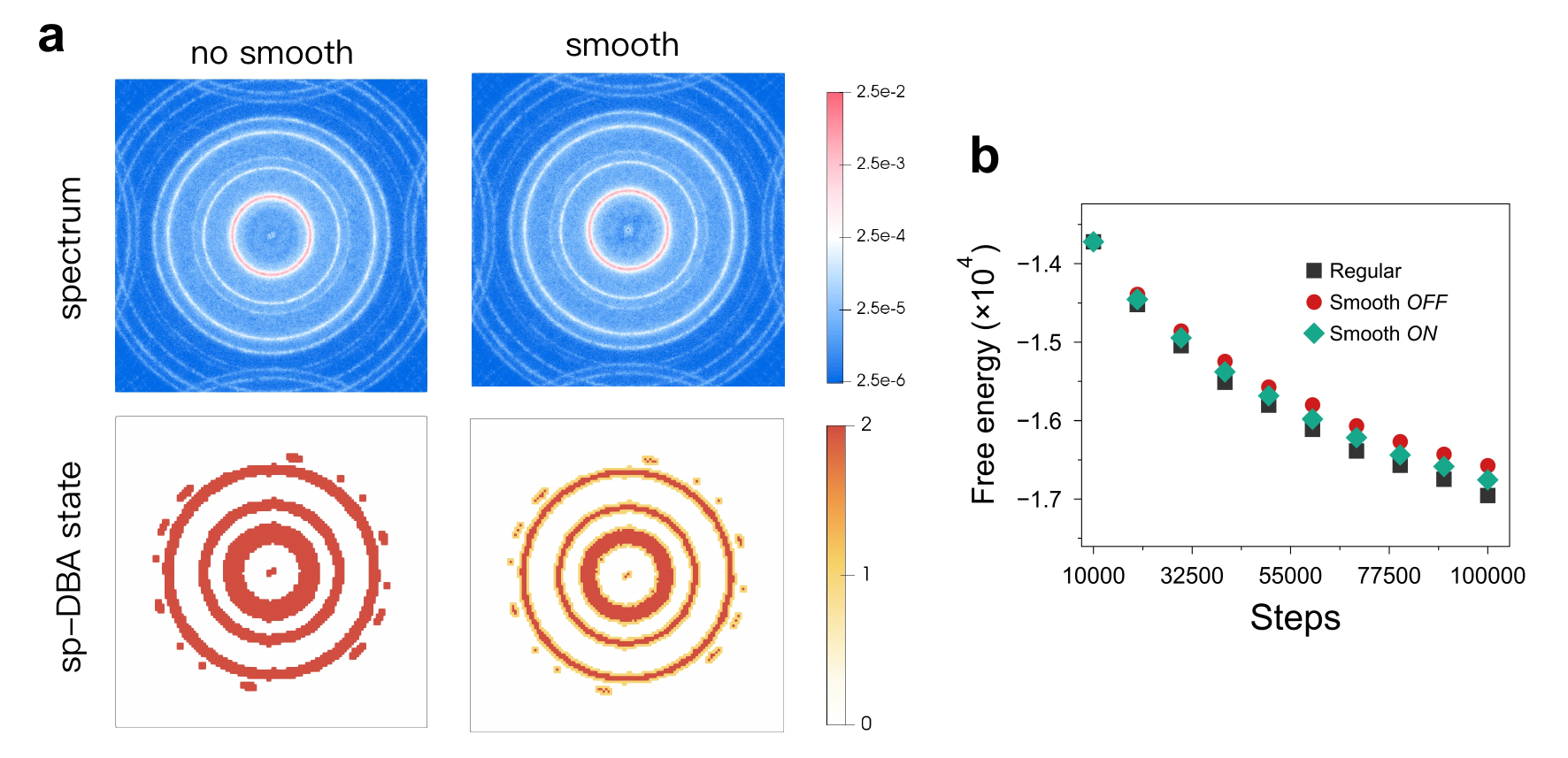}
\caption{Effect of transition smoothing. \textbf{a}, Fourier-amplitude maps and sp-DBA block states without and with smoothing. \textbf{b}, Total free-energy evolution for Regular, sp-DBA without smoothing, and sp-DBA with smoothing.}
\label{fig:s2_smoothing_effect}
\end{figure}

Smoothing leaves the block pattern largely unchanged and brings the free-energy trajectory closer to that of Regular (Fig.~\ref{fig:s2_smoothing_effect}).

\subsection{Neighbor activation in half-spectrum FFT layouts}\label{supp:s1.2}

For real-valued fields, the PFC and RCD solvers store only the nonredundant half of Fourier space in the real-to-complex FFT layout. sp-DBA applies neighbor activation on this stored half-spectrum while preserving the conjugate symmetry required by real-valued fields. If a requested neighbor lies inside the stored half-spectrum, it is activated directly. If the requested neighbor lies in the omitted half, sp-DBA activates its conjugate counterpart using
\begin{equation} \hat{q}(-\bm{k})=\hat{q}^{*}(\bm{k}). \end{equation}

\begin{figure}[H]
\centering
\includegraphics[width=1.0\linewidth]{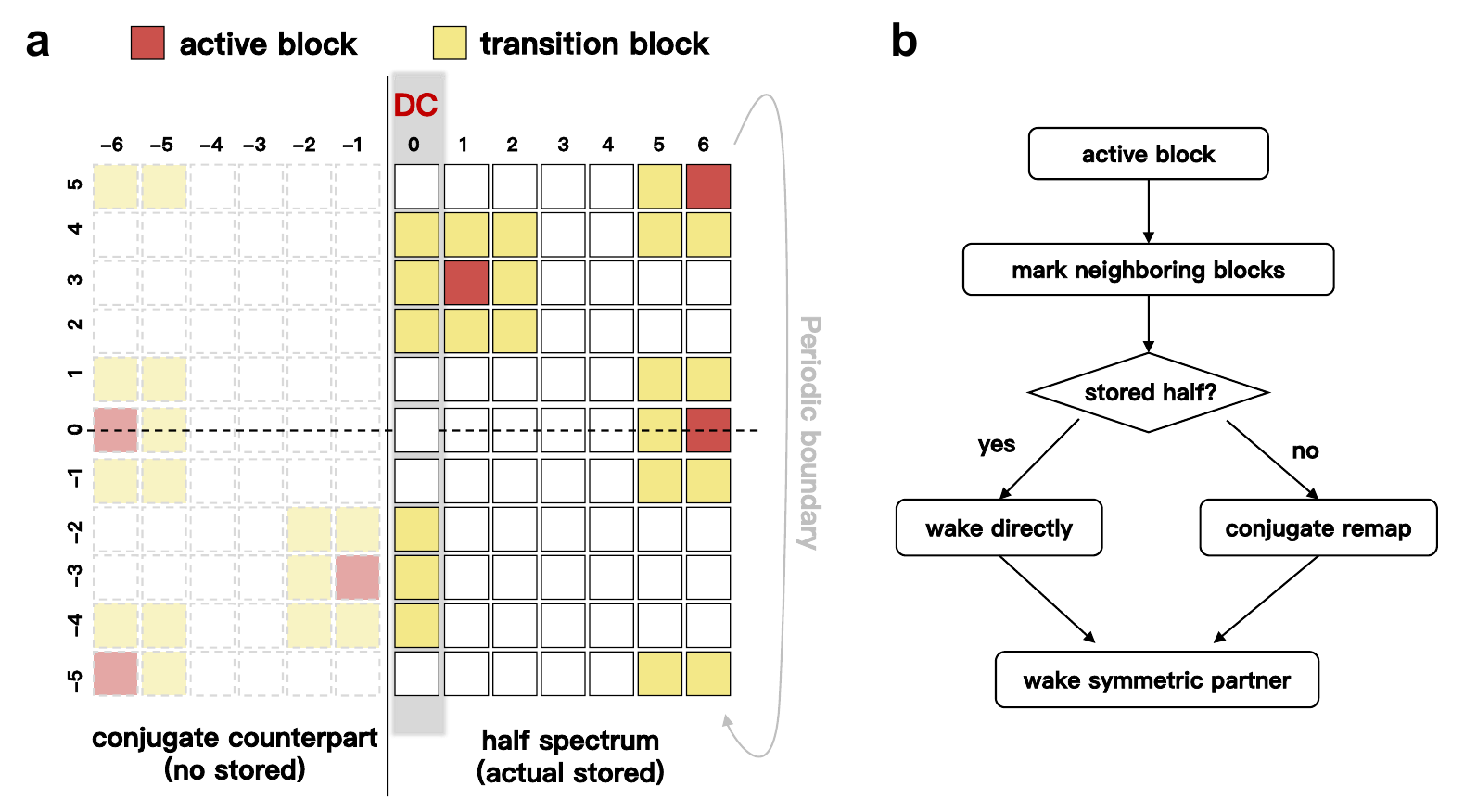}
\caption{Half-spectrum neighbor activation. \textbf{a}, Neighbor activation on the stored half-spectrum. Neighbor blocks outside the stored half are activated through their conjugate counterparts. \textbf{b}, Simplified decision rule. Neighbors inside the stored half are activated directly; neighbors outside it are mapped by conjugate symmetry, and the symmetric partner on the \(k_x=0\) column is activated when required.}
\label{fig:s3_half_spectrum}
\end{figure}

The \(k_x=0\) column is treated as a symmetry boundary. When conjugate remapping reaches this column, the corresponding symmetric partner is also activated. This remapping affects only block-state propagation and does not modify the stored Fourier coefficients.

\subsection{Dynamic threshold control and periodic refresh}\label{supp:s1.3}

\noindent\textbf{Dynamic threshold.}

Dynamic threshold control adjusts the activation threshold to maintain a target average active-block ratio.

At each control interval, the measured average active-block ratio \(r\), including active and transition blocks, is compared with the target ratio \(r_{\mathrm{tar}}\). The threshold is increased when \(r>r_{\mathrm{tar}}\) and decreased when \(r<r_{\mathrm{tar}}\). For PFC, deviations outside a relative deadband \(\eta\) use factors \(1+\eta\) or \(1-\eta\), with factor-of-two corrections when \(r/r_{\mathrm{tar}}>2\) or \(r/r_{\mathrm{tar}}<0.5\).

\begin{figure}[H]
\centering
\includegraphics[width=1.0\linewidth]{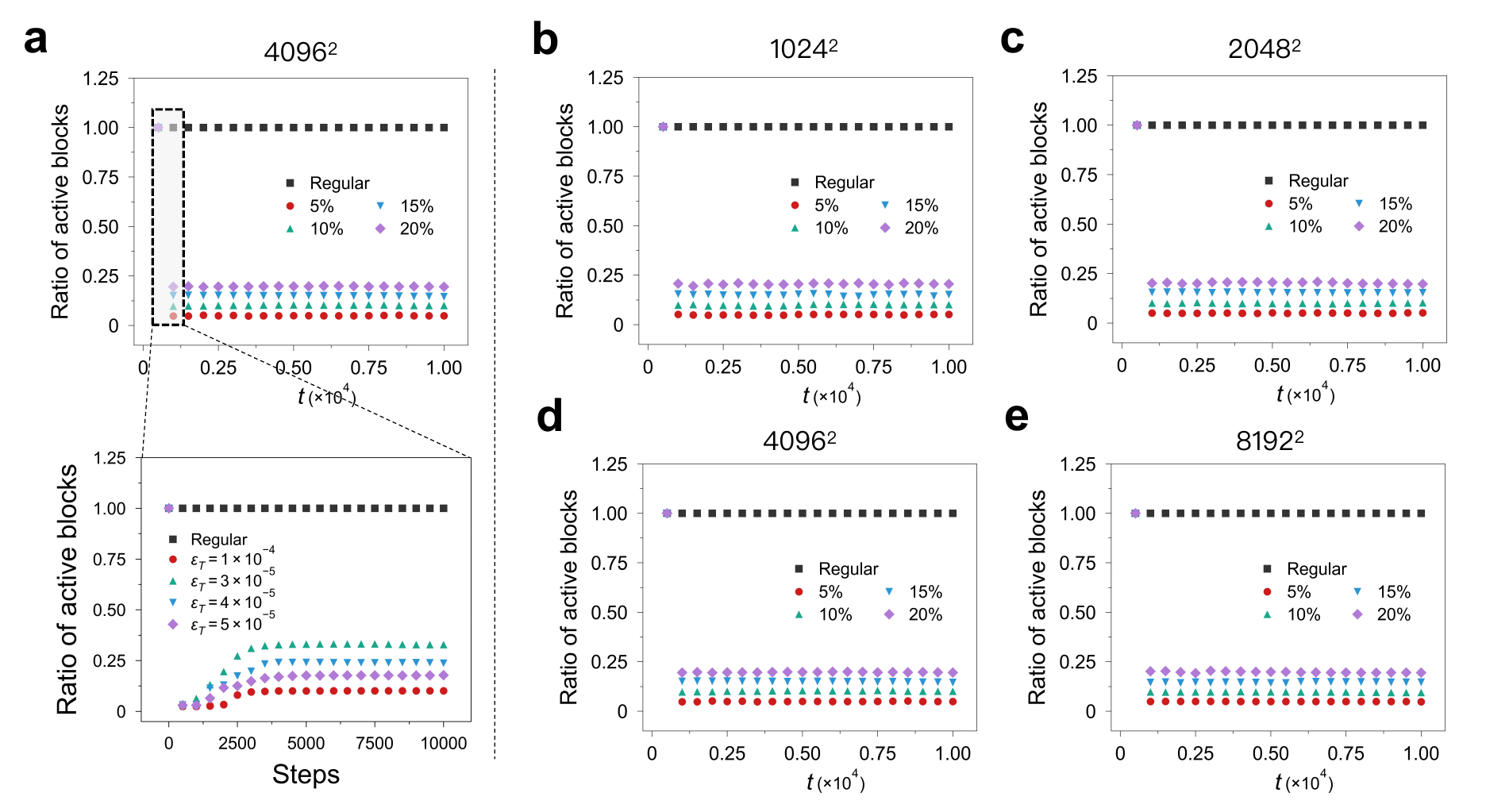}
\caption{Dynamic threshold control in the XPFC application. \textbf{a}, Average active-block ratio on a \(4096^2\) grid. The upper plot shows runs with target ratios of 5\%, 10\%, 15\%, and 20\%; the lower plot shows fixed-threshold behavior during the first 10,000 steps, which are excluded from the reported performance statistics. \textbf{b--e}, Runs with dynamic threshold control on \(1024^2\), \(2048^2\), \(4096^2\), and \(8192^2\) grids. After the initial full-activation stage, the measured active-block ratio stays close to its target.}
\label{fig:s4_dynamic_threshold}
\end{figure}

In the reported PFC runs, control begins after the initial 10,000-step full-activation stage, and each ratio check is followed by one full-activation step.

\noindent\textbf{Periodic refresh.}

Periodic refresh activates all blocks every $F_{\mathrm{refresh}}$ steps to limit long-time drift. Between refresh events, blocks selected by the activation rule are updated at the base time step; for example, $\Delta t=0.01$ and $F_{\mathrm{refresh}}=50$ give a refresh interval of 0.5 time units.

\begin{figure}[H]
\centering
\includegraphics[width=1.0\linewidth]{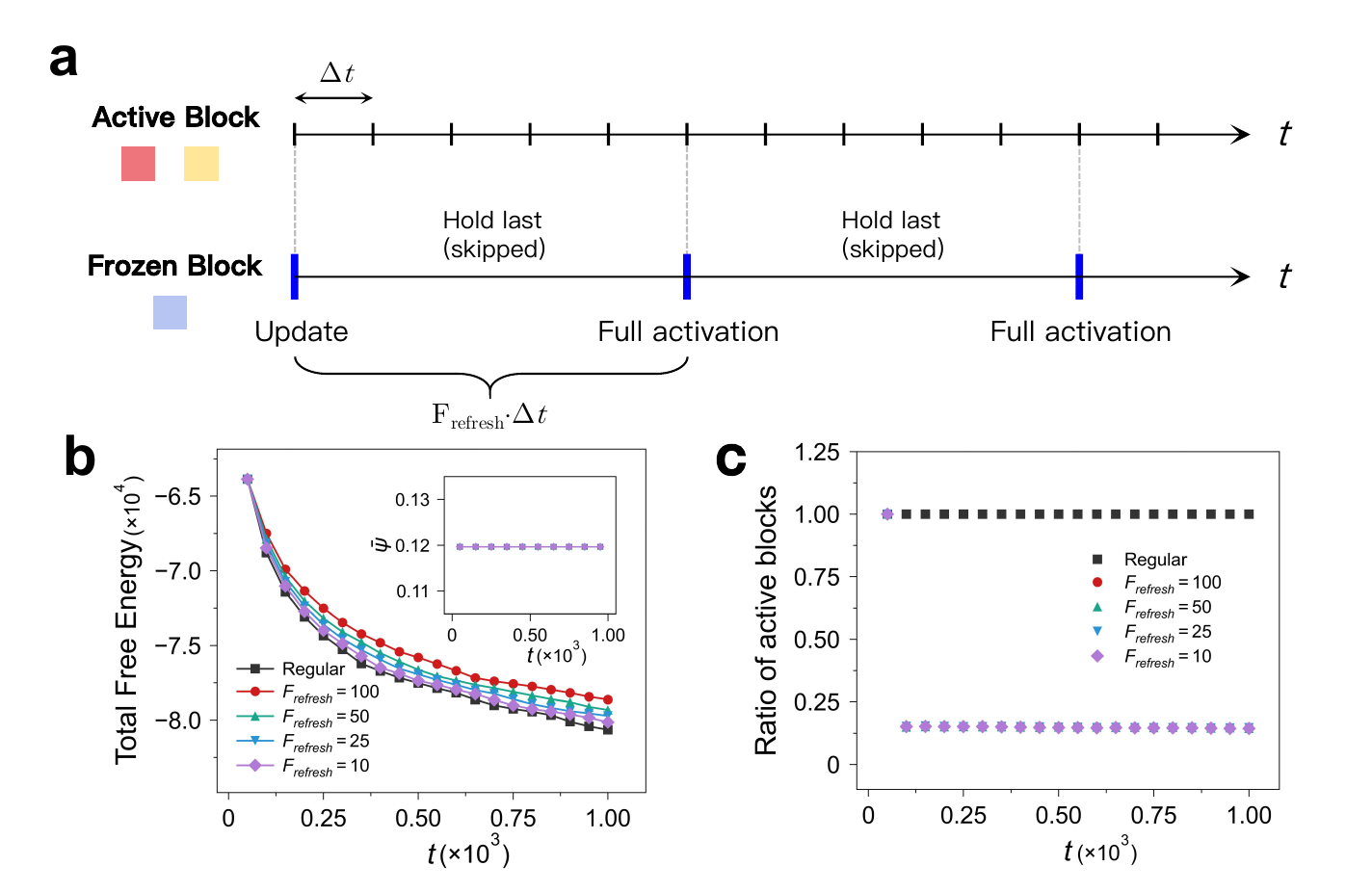}
\caption{Periodic refresh. \textbf{a}, Active blocks are updated at each step $\Delta t$. Frozen blocks retain their previous state between local reactivation events or full-activation events separated by $F_{\mathrm{refresh}}\Delta t$. \textbf{b}, Total free-energy evolution for Regular and sp-DBA with different refresh intervals. The inset shows the mean density \(\bar{\psi}\). \textbf{c}, Average active-block ratio for the same refresh intervals.}
\label{fig:s5_refresh_interval}
\end{figure}

Shorter refresh intervals bring the free-energy curves closer to the Regular trajectory, while the average active-block ratio returns to its low baseline after each refresh (Fig.~\ref{fig:s5_refresh_interval}). The mean density is unchanged over the tested intervals.

% ======================================================================
\section{sp-DBA integration with transform-domain solvers}
\label{sec:solver_integration}
% ======================================================================

\begin{figure}[H]
\centering
\includegraphics[width=1.0\linewidth]{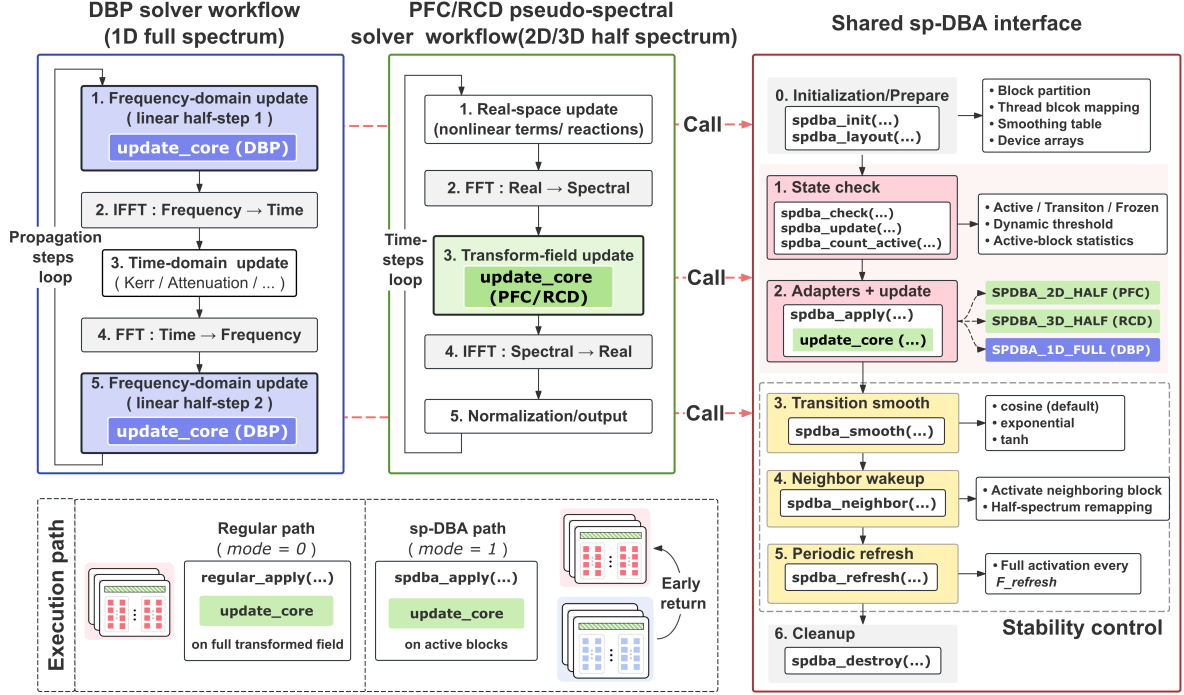}
\caption{Integration of sp-DBA into the three solver workflows. DBP connects sp-DBA to the two frequency-domain linear half-steps of the one-dimensional full-spectrum split-step solver, while PFC and RCD connect it to the transformed-field update of two- and three-dimensional pseudo-spectral solvers. Regular and sp-DBA call the same model-specific \texttt{update\_core}; sp-DBA controls which blocks execute it and how transition increments are weighted. The shared interface initializes block states, dispatches the update, applies transition handling, updates neighboring states, performs periodic refresh, and releases device arrays.}
\label{fig:s6_solver_integration}
\end{figure}

Figure~\ref{fig:s6_solver_integration} summarizes the shared sp-DBA interface used in the three solvers.

\subsection{Integration procedure}\label{supp:s2.1}

Each solver exposes its transformed-field update as \texttt{update\_core}. Regular calls this routine over the full transformed field, whereas sp-DBA calls it only for active and transition blocks. Frozen blocks return early, and transition blocks use the weighted increment defined in Supplementary Section~\ref{sec:core_mechanisms}.

The solver invokes \texttt{spdba\_apply(...)} at this stage. With \texttt{mode=0}, it dispatches the full update; with \texttt{mode=1}, it checks block states, calls \texttt{update\_core} where required, updates neighboring states, and applies threshold or refresh control.

Each model adapter supplies the layout, indexing, launch configuration, and model parameters required by the shared interface. It binds the model-specific \texttt{update\_core} to the generic kernel for the one-dimensional full-spectrum DBP layout and the two- and three-dimensional half-spectrum PFC and RCD layouts.

\begin{framed}
\noindent\textbf{Pseudocode for integration of sp-DBA into a solver}
\begin{verbatim}
readInput(...)
allocate model fields
construct spDBA parameters, state arrays, and model adapter
if (params.mode == 1) spdba_init(...)

for each simulation step
    model_pre_update(...)       // model terms, transforms, preprocessing
    construct spDBA_step(...)
    spdba_apply(...)            // mode=0 full update; mode=1 sp-DBA
    model_post_update(...)      // inverse transforms, normalization, output

if (params.mode == 1) spdba_destroy(...)
\end{verbatim}
\end{framed}

The generic kernels construct block-local indices and call the model-provided \texttt{update\_core} for the \texttt{SPDBA\_1D\_FULL}, \texttt{SPDBA\_2D\_HALF}, and \texttt{SPDBA\_3D\_HALF} layouts.

\subsection{Implementation overhead}\label{supp:s2.2}

Implementation overhead is measured in nonblank lines of code (LOC), following Ref.~\cite{2025_dba}, with separate counts for the shared core, model adapter, and solver integration. The shared \texttt{spDBA/} core contains 696 LOC and implements block partitioning, state management, transition smoothing, neighbor activation, threshold control, and refresh; it is excluded from the counts and percentages for each solver. Simple changes to each solver include exposing \texttt{update\_core}, binding the sp-DBA state, reading parameters, connecting diagnostic output, and calling the shared interface. Input/output code, scripts, documentation, tests, generated files, and historical files are excluded.

\begin{table}[H]
\centering
\small
\caption{Implementation overhead for integrating sp-DBA into the three solvers.}
\label{tab:implementation_overhead}
\setlength{\tabcolsep}{3pt}
\begin{tabular}{lrrrrrrr}
\toprule
Application & Regular core & Solver LOC & Solver \% & Adapter & Combined LOC & Combined \% & API calls \\
\midrule
PFC & 2162 & 178 & 8.23\% & 117 & 295 & 13.64\% & 9 \\
RCD & 1284 & 211 & 16.43\% & 145 & 356 & 27.73\% & 11 \\
DBP & 1513 & 159 & 10.51\% & 97 & 256 & 16.92\% & 19 \\
\bottomrule
\end{tabular}
\end{table}

The simple changes add 159--211 nonblank LOC, corresponding to about 8--16\% of each Regular solver core. Including the model adapters increases the total to 256--356 LOC, or about 13--28\% of the original core sizes. RCD has the largest percentage because its integration includes a three-dimensional half-spectrum layout, multicomponent fields, eigendecomposition, legacy update paths, and shared-memory launch configuration.

\subsection{Comparison with adaptive spectral and wavelet methods}\label{supp:s2.3}
\label{sec:si_adaptive_workflow_comparison}

Table~\ref{tab:adaptive_workflow_comparison} compares how representative adaptive spectral, wavelet, and sp-DBA workflows introduce adaptivity, focusing on changes to the numerical representation, data structures, and solver implementation. Runtime and accuracy are not compared across methods.

Representative adaptive spectral methods modify the approximation space during the calculation. Frequency-dependent $p$-adaptivity adjusts the expansion order and reconstructs the solution on a new set of collocation points, while adaptive Hermite methods update basis parameters and map the numerical state between successive approximation spaces~\cite{2021_p_adaptive,2023_adp_sp2}.

Adaptive wavelet workflows introduce a hierarchy of resolution levels. Wavelet-enhanced FFT methods refine the computational grid and reformulate the Fourier-space operator for the wavelet discretization, while parallel implementations use hierarchical storage, refinement, coarsening, reconstruction or ghost regions, and, in distributed calculations, block migration and dynamic load balancing~\cite{2023_adp_wavelet_layerFFT,2015_pawcm,2022_wavelet_pod}.

sp-DBA instead retains the transformed representation, regular array layout, transform operations, and model-specific update core, while adding block state arrays and activation control around the existing update. Table~\ref{tab:adaptive_workflow_comparison} summarizes the contrast.

\begin{table}[H]
\centering
\small
\caption{Representative implementation paths for adaptive spectral and wavelet workflows and sp-DBA.}
\label{tab:adaptive_workflow_comparison}
\renewcommand{\arraystretch}{1.18}
\begin{tabular}{
>{\raggedright\arraybackslash}p{0.23\linewidth}
>{\raggedright\arraybackslash}p{0.25\linewidth}
>{\raggedright\arraybackslash}p{0.44\linewidth}}
\toprule
Adaptive strategy & Quantity changed during the calculation & Additional implementation path \\
\midrule
Adaptive spectral basis & Expansion order or basis parameters & Reconstruct or project the numerical state between successive approximation spaces \\
Wavelet and multiresolution workflow & Active resolution levels, grid points or wavelet coefficients & Maintain hierarchical grid data and manage refinement, coarsening, reconstruction or ghost regions, and, where required, migration and load balancing \\
sp-DBA & Blocks that execute the existing transformed-field update & Maintain block states and activation control around the retained update core; the representation, regular data layout, and transform operations remain in place \\
\bottomrule
\end{tabular}
\end{table}

Implementation requirements vary among adaptive spectral and wavelet methods. Table~\ref{tab:implementation_overhead} reports the added code for the three sp-DBA integrations and is not a cross-method LOC comparison.

% ============================================================
\section{XPFC polycrystalline coarsening}
\label{sec:si_pfc}
% ============================================================

The XPFC tests use a scalar pseudo-spectral model for polycrystalline coarsening.

\subsection{Governing equations}\label{supp:s3.1}

We use a dimensionless XPFC formulation~\cite{2010_XPFC_prl,2023_tmd_xia}. The order parameter $\psi(\bm{r})$ describes deviations from a reference density and evolves according to a free energy with local polynomial terms and a nonlocal two-point correlation term,
\begin{equation} F[\psi] = \int d\bm{r} \left[ \frac12\psi^2-\frac{\omega}{6}\psi^3+\frac{u}{12}\psi^4 \right] -\frac12 \iint d\bm{r}\,d\bm{r}'\, \psi(\bm{r}) C_2(|\bm{r}-\bm{r}'|)\psi(\bm{r}'). \end{equation}
The local terms define the bulk free energy and stabilize the density amplitude, while the nonlocal term favors density modulations selected by the correlation kernel $C_2$. The conserved dynamics follow the Cahn--Hilliard form~\cite{1958_cahn_hilliard},
\begin{equation} \frac{\partial \psi}{\partial t} = \nabla^2 \frac{\delta F}{\delta \psi}, \end{equation}
which gives
\begin{equation} \frac{\partial \psi}{\partial t}= \nabla^2 \left[ \psi -\int d\bm{r}'\, C_2(|\bm{r}-\bm{r}'|)\psi(\bm{r}') -\frac{\omega}{2}\psi^2 +\frac{u}{3}\psi^3 \right]. \end{equation}
The outer Laplacian enforces conservation of the mean density.

In Fourier space, the convolution with $C_2$ becomes multiplication by $\hat{C}_2(k)$. For the two-dimensional hexagonal XPFC setting used here, the reciprocal-space kernel is represented by a single Gaussian peak,
\begin{equation} \hat{C}_2(k) = \exp\!\left[-\frac{(k-q_0)^2}{2\alpha^2}\right] \exp\!\left[-\frac{\sigma^2 q_0^2}{2\rho\beta}\right], \end{equation}
with $\rho=\beta=1$ in the production coarsening runs. The first exponential defines the structural peak around the preferred wavenumber $q_0$, and the second exponential acts as a Debye--Waller attenuation factor.

The pseudo-spectral implementation evaluates the local nonlinear terms in real space and the correlation term in Fourier space. In polycrystalline states, different grain orientations broaden the Bragg peaks into ring-like Fourier-amplitude regions, which are used to assign the PFC block states.

\subsection{Numerical scheme and simulation configuration}\label{supp:s3.2}

\noindent\textbf{Time-stepping method.}

The simulations are advanced in Fourier space with the fourth-order multistep exponential time-differencing scheme (ETD4) of Cox and Matthews~\cite{2002_etd}. ETD4 advances the linear term through the exponential propagator \(e^{\mathcal{L}\Delta t}\) and evaluates the nonlinear contribution using its current value and three previous Fourier-space values. Writing the Fourier-space evolution as
\begin{equation}
\frac{d\widehat{\psi}}{dt}=\mathcal{L}(\bm{k})\widehat{\psi}+\widehat{\mathcal{N}(\psi)},
\end{equation}
the update is
\begin{equation}
\begin{aligned}
\widehat{\psi}^{\,n+1}={}&e^{z}\widehat{\psi}^{\,n}+\widehat{N}^{\,n}\Phi_1+\left(\widehat{N}^{\,n}-\widehat{N}^{\,n-1}\right)\Phi_2+\left(\widehat{N}^{\,n}-2\widehat{N}^{\,n-1}+\widehat{N}^{\,n-2}\right)\Phi_3\\
&+\left(\widehat{N}^{\,n}-3\widehat{N}^{\,n-1}+3\widehat{N}^{\,n-2}-\widehat{N}^{\,n-3}\right)\Phi_4,
\end{aligned}
\end{equation}
where \(\widehat{N}^{\,n}=\widehat{\mathcal{N}(\psi^n)}\), \(z=\mathcal{L}(\bm{k})\Delta t\), and
\begin{equation}
\begin{aligned}
\Phi_1&=\Delta t\,\phi_1(z),\\
\Phi_2&=\Delta t\,\phi_2(z),\\
\Phi_3&=\Delta t\left[\frac{1}{2}\phi_2(z)+\phi_3(z)\right],\\
\Phi_4&=\Delta t\left[\frac{1}{3}\phi_2(z)+\phi_3(z)+\phi_4(z)\right].
\end{aligned}
\end{equation}
The coefficient functions are defined by
\begin{equation}
\phi_\ell(z)=\sum_{j=0}^{\infty}\frac{z^j}{(\ell+j)!},
\qquad\ell=1,\ldots,4 .
\end{equation}
For small \(z\), the coefficient functions are evaluated using their Taylor expansions to avoid cancellation and apparent division by zero~\cite{2002_etd,2005_etd4_Talyor}. In the conserved PFC implementation, the outer Fourier-space Laplacian is included in \(\widehat{\mathcal{N}(\psi)}\), and the corresponding factor \(-k^2\Delta t\) is incorporated into the precomputed nonlinear-update coefficients. At initialization, the three nonlinear-history arrays are set to the initial nonlinear term and shifted after each time step.

\noindent\textbf{Simulation and initialization parameters.}

The XPFC simulations are performed on a two-dimensional periodic square domain. Performance tests use grid sizes from $512^2$ to $16384^2$. The layered main-text visualization uses a $2048^2$ production run, whereas the main-text GSD and free-energy comparisons use $8192^2$ production runs. The representative production configuration uses $\Delta x=\Delta y=0.7854$, $\Delta t=0.01$, and $1,000,000$ time steps, corresponding to a final dimensionless time of $t=10,000$. The XPFC evolution parameters are $q_0=1.0$, $\sigma=0.02$, $\alpha=0.1$, $\omega=1.0$, and $u=1.0$. Here, $q_0$, $\sigma$, and $\alpha$ parameterize the reciprocal-space correlation kernel, while $\omega$ and $u$ define the local nonlinear terms in the free energy. The polycrystalline initial condition is constructed around a mean density $\psi_c=0.12$ by adding randomly oriented crystalline seeds with density modulation amplitude $A_h=0.25$. The seed centers and orientations are generated from fixed random seeds and are smoothed near seed boundaries before the initial growth stage.

For the sp-DBA runs, the Fourier block size is $B=16$. The initial threshold is $5\times10^{-5}$, boundary-aware transition smoothing and dynamic threshold control are enabled, the adjustment factor is 0.05, and the control interval is 50 steps. The target average active-block ratio ranges from 5\% to 20\%. During the first 10,000 steps, all Fourier blocks are kept active to avoid premature freezing during the rapidly evolving initial transient. The late-stage analysis begins after this full-activation period.

\begin{figure}[H]
\centering
\includegraphics[width=0.85\linewidth]{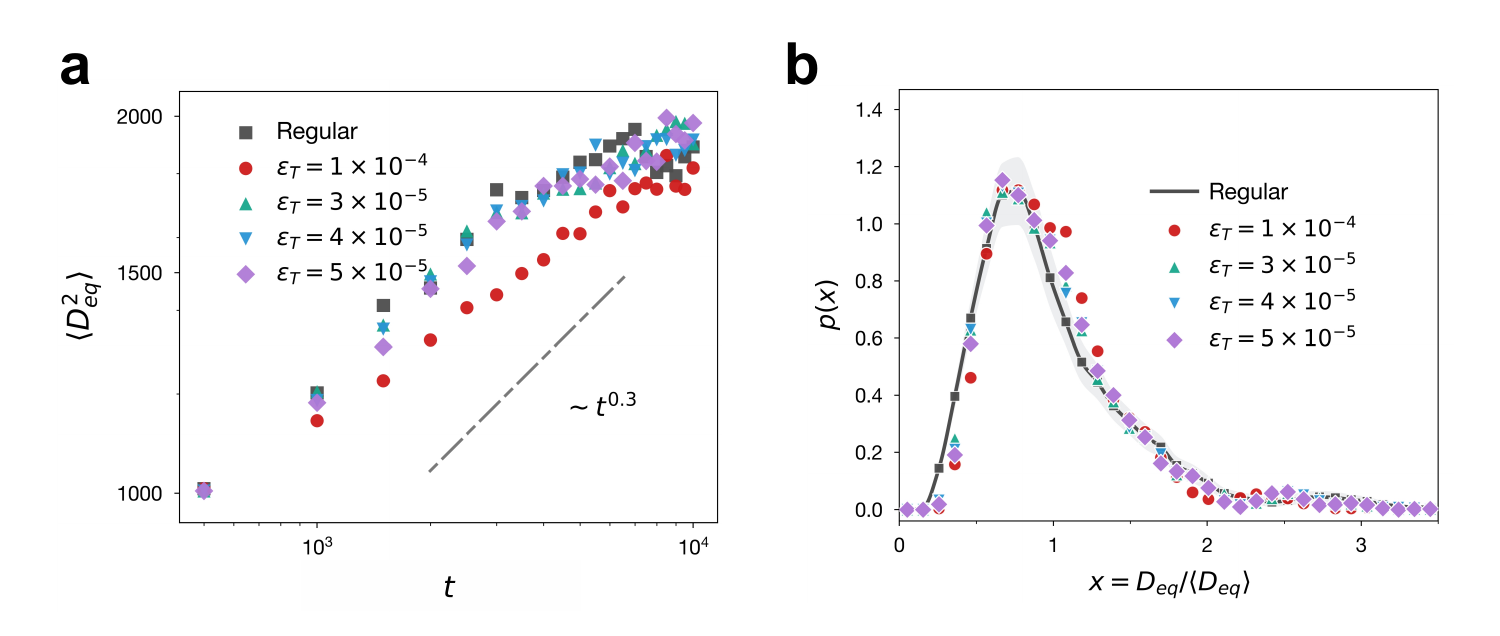}
\caption{Fixed-threshold sensitivity in the XPFC application. \textbf{a}, Mean squared equivalent grain diameter $\langle D_{\mathrm{eq}}^2\rangle$ as a function of sampled coarsening time for Regular and sp-DBA with fixed threshold values $\epsilon_T$ on a $4096^2$ grid. \textbf{b}, Normalized grain-size distribution $p(x)$ with $x=D_{\mathrm{eq}}/\langle D_{\mathrm{eq}}\rangle$ for the same fixed threshold settings. The gray band indicates the Regular result range.}
\label{fig:s_pfc_fixed_threshold}
\end{figure}

Fixed-threshold sensitivity is also tested on a $4096^2$ grid using the same simulation and sp-DBA configuration. Only the threshold policy changes, with fixed values $\epsilon_T=1\times10^{-4}$, $3\times10^{-5}$, $4\times10^{-5}$, and $5\times10^{-5}$. Across these values, the normalized grain-size distribution and grain-growth trend agree with Regular (Fig.~\ref{fig:s_pfc_fixed_threshold}). The main text uses dynamic threshold control because it keeps the average active-block ratio more stable across coarsening stages and grid sizes.

\noindent\textbf{Layered correspondence.}

\begin{figure}[H]
\centering
\includegraphics[width=1.0\linewidth]{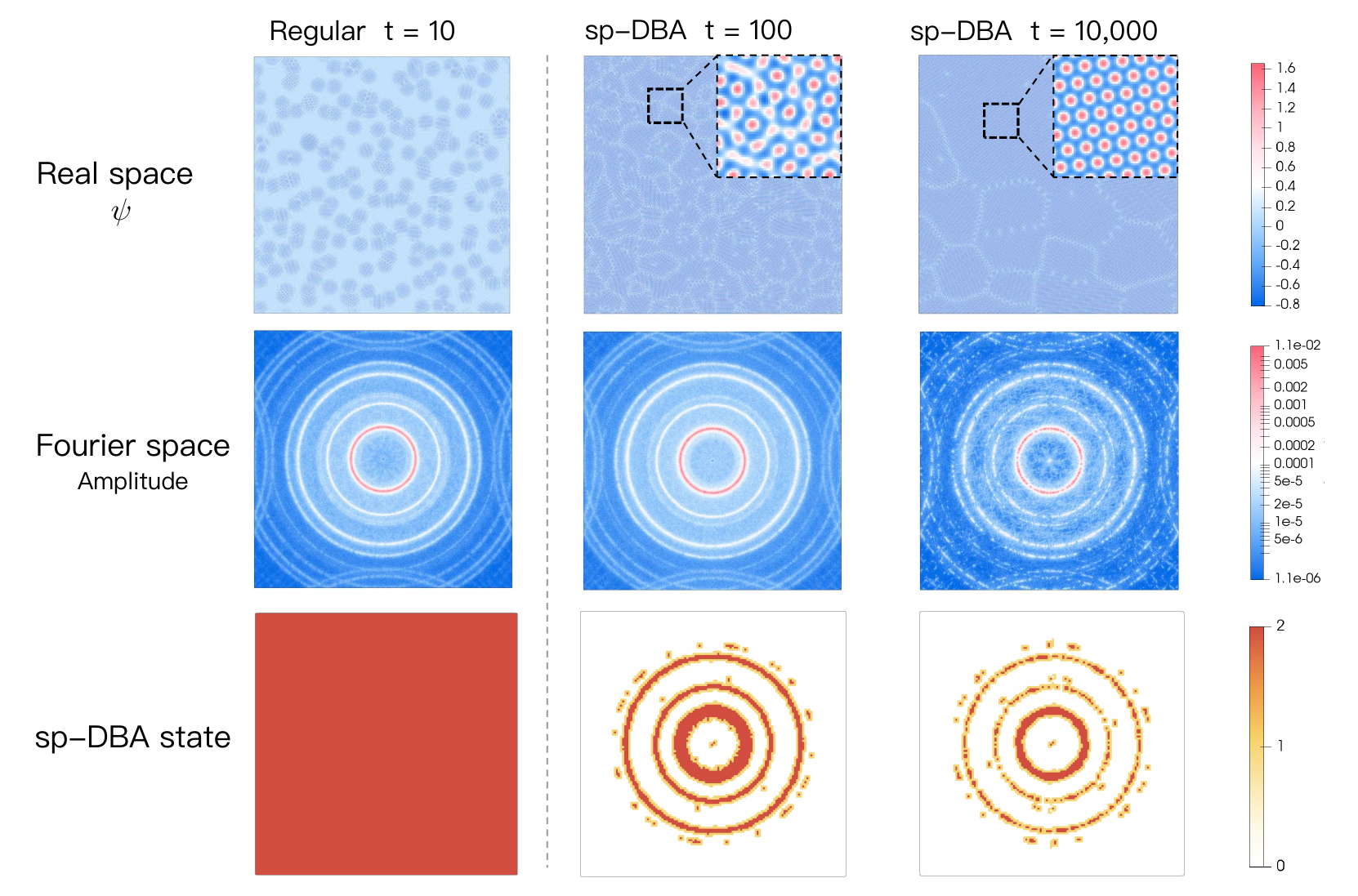}
\caption{Layered XPFC coarsening under sp-DBA on a $2048\times2048$ grid. The top, middle, and bottom rows show the real-space density field $\psi$, Fourier amplitude, and sp-DBA block state, respectively. The average active-block ratio is about 15\%.}
\label{fig:s6_pfc_three_layer}
\end{figure}

The real-space density evolves through grain-boundary migration and defect elimination, while the Fourier amplitude and active blocks are concentrated near the dominant Bragg ring.

\subsection{Grain reconstruction and system-size comparison}\label{supp:s3.3}

Grain statistics are reconstructed from detected density peaks following Ref.~\cite{2018_grain_exact}. Because the labels contain post-processing noise, the late-stage scaling plots are treated as qualitative agreement checks~\cite{2014_gsd2}.

\noindent\textbf{Extraction procedure.}

The continuous density field is first converted into an atom-like point set by detecting local density maxima under periodic boundary conditions. For each detected peak, a first-neighbor shell is constructed within a cutoff radius, and a local bond-orientational descriptor is evaluated. For the two-dimensional hexagonal lattice, the orientation is represented by the phase of the sixfold bond-order parameter.

Atoms with sufficiently ordered local environments are selected as grain cores, while atoms with weak orientational order or strong distortion are treated as boundary or defect candidates. Neighboring core atoms are merged into the same grain when their misorientation remains below the prescribed large-angle boundary threshold. Excluded boundary atoms are then reassigned by neighborhood voting and proximity checks.

Very small fragments, enclosed islands and low-misorientation artifacts are removed or merged. The cleaned labels and a local atomic-area approximation are then used to evaluate grain area, equivalent diameter, GSD and coarsening statistics in a form comparable with standard grain-growth analysis~\cite{2014_gsd2}.

\noindent\textbf{Extraction settings and cross-size comparison.}

The same reconstruction workflow is applied to the sampled OVITO dump frames from the Regular run and from sp-DBA runs with average active-block ratios of 5\%, 10\%, 15\%, and 20\%. The fixed reconstruction parameters are $r_{\mathrm{cut}}=1.18$, $\theta_{\mathrm{cut}}=3.5^\circ$, $\psi_{\mathrm{keep}}=0.55$, and a minimum grain core size of 80 atoms. Boundary atoms are reassigned using six voting passes with an assignment radius factor of 1.3, and small-fragment cleanup uses a merge factor of 3.0.

For the normalized GSD, grain diameters are rescaled frame by frame as $x=D_{\mathrm{eq}}/\langle D_{\mathrm{eq}}\rangle$, and the steady distribution is constructed from the last 50\% of sampled frames. The histogram uses 22, 28, 34, and 40 bins for $1024^2$, $2048^2$, $4096^2$, and $8192^2$, respectively, with the plotted range limited to $0<x<3.5$. The supplementary scaling plots use sampled coarsening time $t$ and show the common guide line $\langle D_{\mathrm{eq}}^{2}\rangle\sim t^{0.3}$ for visual comparison.

\begin{figure}[H]
\centering
\includegraphics[width=1.0\linewidth]{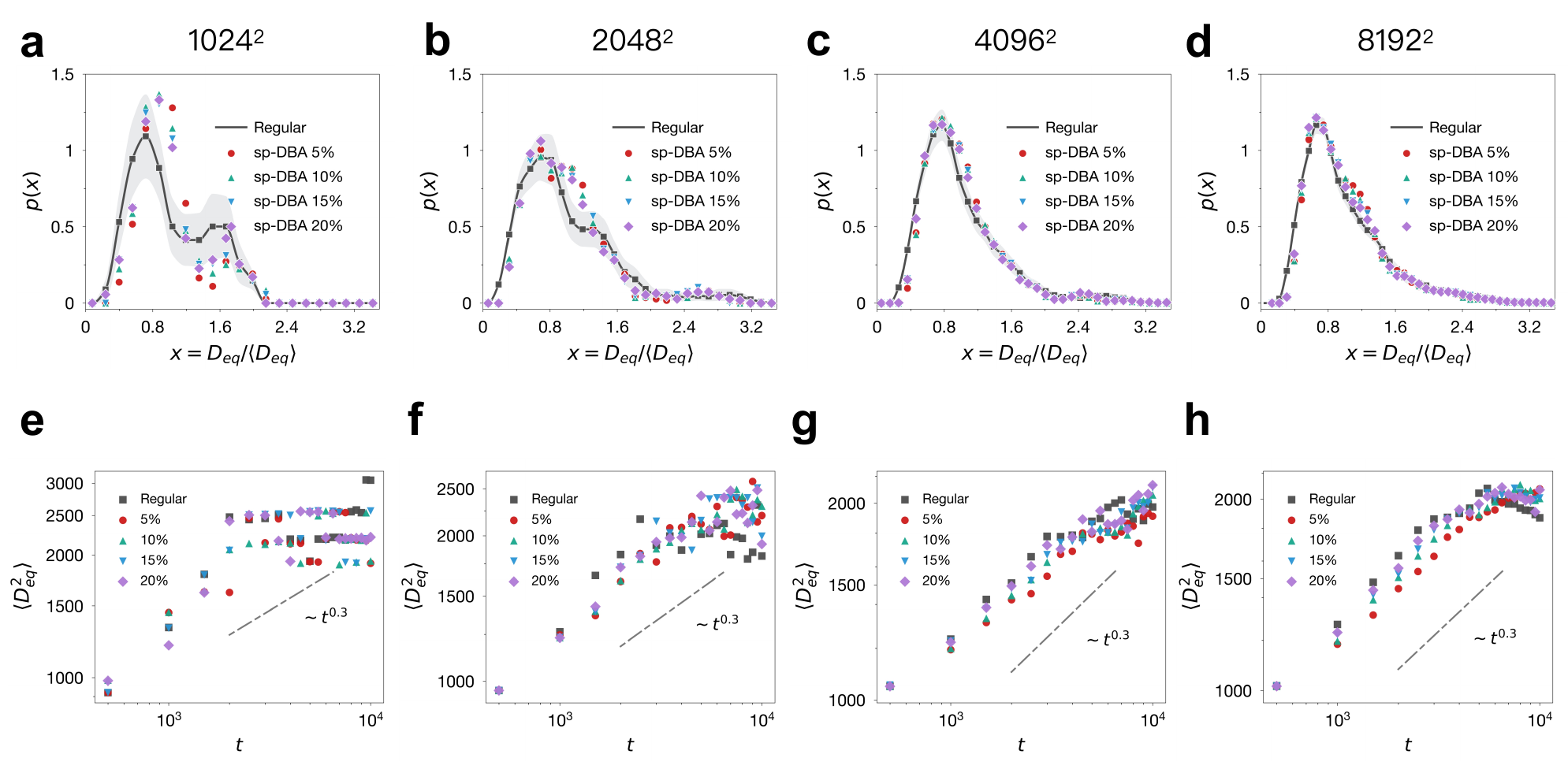}
\caption{Supplementary GSD and late-stage scaling comparisons across XPFC sizes. \textbf{a--d}, Steady normalized GSD comparisons for the $1024^2$, $2048^2$, $4096^2$, and $8192^2$ production runs. \textbf{e--h}, Corresponding sampled late-stage scaling plots for the same four sizes, shown with the common guide line $\langle D_{\mathrm{eq}}^{2}\rangle\sim t^{0.3}$.}
\label{fig:s7_pfc_gsd_scaling_sizes}
\end{figure}

The GSD collapse becomes clearer with increasing system size, with the largest systems giving the smoothest late-stage distributions (Fig.~\ref{fig:s7_pfc_gsd_scaling_sizes}).

% ============================================================
% Keep S4--S5 at section level in the global Contents.
\addtocontents{toc}{\protect\setcounter{tocdepth}{1}}
\section{Reaction--cross-diffusion pattern formation}
\label{sec:si_rcd}
% ============================================================

The RCD tests use a constant-coefficient multicomponent model.

\subsection{Governing equations}\label{supp:s4.1}

The general multicomponent reaction--cross-diffusion system is written as~\cite{2007_cd_review,2014_LV_RCD1}
\begin{equation} \frac{\partial \bm{u}}{\partial t} = \nabla\cdot\left(\mathbf{D}(\bm{u})\nabla \bm{u}\right) + \mathbf{R}(\bm{u}), \end{equation}
where \(\bm{u}=(u_1,\ldots,u_M)^T\) is the concentration vector, \(\mathbf{D}(\bm{u})\) is the diffusion matrix, and \(\mathbf{R}(\bm{u})\) is the reaction term. In the reported tests, the model is reduced to the constant-coefficient form
\begin{equation} \frac{\partial \bm{u}}{\partial t} = \mathbf{D}\nabla^2\bm{u} + \mathbf{R}(\bm{u}). \end{equation}

The diffusion matrix is chosen as a symmetric circulant cross-diffusion matrix over the component index,
\begin{equation} \mathbf{D}= \begin{pmatrix} D_{\mathrm{diag}}+2D_{\mathrm{cross}} & -D_{\mathrm{cross}} & 0 & \cdots & -D_{\mathrm{cross}} \\ -D_{\mathrm{cross}} & D_{\mathrm{diag}}+2D_{\mathrm{cross}} & -D_{\mathrm{cross}} & \cdots & 0 \\ 0 & -D_{\mathrm{cross}} & D_{\mathrm{diag}}+2D_{\mathrm{cross}} & \cdots & \vdots \\ \vdots & \vdots & \vdots & \ddots & -D_{\mathrm{cross}} \\ -D_{\mathrm{cross}} & 0 & \cdots & -D_{\mathrm{cross}} & D_{\mathrm{diag}}+2D_{\mathrm{cross}} \end{pmatrix}. \end{equation}

For \(M=2\), the two periodic neighbors coincide, and the single off-diagonal entry is \(-D_{\mathrm{cross}}\).
Each component is coupled to its two neighboring components under periodic indexing. Because \(\mathbf{D}\) is constant, the optional eigendecomposition described in Supplementary Section~\ref{sec:si_rcd_eigen} is computed once before time stepping~\cite{2020_matrix_rd}.

The nonlinear reaction term is taken in generalized Lotka--Volterra form~\cite{2007_cd_review,2014_LV_RCD1,2009_LV_RCD2_topo},
\begin{equation} R_i(\bm{u})= u_i \left( r_i-a_i u_i-\sum_{j\neq i}\alpha_{ij}u_j \right). \end{equation}
Here, \(r_i\) is the intrinsic growth rate, \(a_i\) is the self-limitation coefficient, and \(\alpha_{ij}\) describes the interaction from component \(j\) to component \(i\). In the reported tests, \(r_i=1\), \(a_i=1\), \(\alpha_{i,i+1}=\alpha_{i,i-1}=\beta+\alpha\), and \(\alpha_{ij}=\beta\) for all other off-diagonal entries, again with periodic indexing over components.

\subsection{Numerical scheme and simulation configuration}\label{supp:s4.2}

\noindent\textbf{Time stepping method.}

Time integration uses a semi-implicit Fourier update. The linear cross-diffusion term is treated implicitly, and the nonlinear reaction term is evaluated explicitly. For each stored Fourier mode \(\bm{k}\),
\begin{equation} \hat{\bm{u}}^{\,n+1}(\bm{k}) = \left(\mathbf{I}+\Delta t\,k^2\mathbf{D}\right)^{-1} \left[ \hat{\bm{u}}^{\,n}(\bm{k}) + \Delta t\,\widehat{\mathbf{R}}(\bm{u}^{\,n})(\bm{k}) \right]. \end{equation}
Thus, in Regular, each stored Fourier mode requires an \(M\times M\) matrix operation.

\noindent\textbf{Simulation parameters.}

The tested configurations use periodic three-dimensional domains with spatial resolutions of \(128^3\) and \(256^3\), while the number of components is varied from \(M=2\) to \(M=64\). The spatial spacing is \(\Delta x=0.5\), the time step is \(\Delta t=0.01\), and the total number of steps is 20,000. The constant diffusion matrix uses \(D_{\mathrm{diag}}=0.5\) and \(D_{\mathrm{cross}}=0.05\). The generalized Lotka--Volterra reaction uses \(\alpha=1.0\) for the enhanced neighboring component interaction and \(\beta=0.3\) for the remaining intercomponent competition. The initial condition is a nearly homogeneous state with random perturbation amplitude 0.1.

For the sp-DBA runs, the block size is \(B=4\) for \(128^3\) and \(B=8\) for \(256^3\), the initial threshold is \(10^{-9}\), and dynamic threshold control is enabled. The adjustment factor is 0.1, the control interval is 50 steps, and the target average active-block ratio ranges from 3\% to 20\%.

For RCD block selection, the activation criterion is computed from all components rather than from the single component shown in visualizations. At each stored Fourier mode \(\bm{k}\), the implementation uses the component RMS Fourier amplitude
\begin{equation} \mathcal{A}(\bm{k}) = \left( \frac{1}{M} \sum_{m=1}^{M} |\hat{u}_m(\bm{k})|^2 \right)^{1/2}. \end{equation}
A Fourier block \(B_j\) is activated when at least one stored mode satisfies \(\mathcal{A}(\bm{k})>\varepsilon_{\mathrm{RCD}}\), and the resulting state is shared by all \(M\) components in that block. The eigendecomposition of \(\mathbf{D}\) is used only within the update core and does not enter the activation criterion.

\noindent\textbf{Layered correspondence.}

\begin{figure}[H]
\centering
\includegraphics[width=1.0\linewidth]{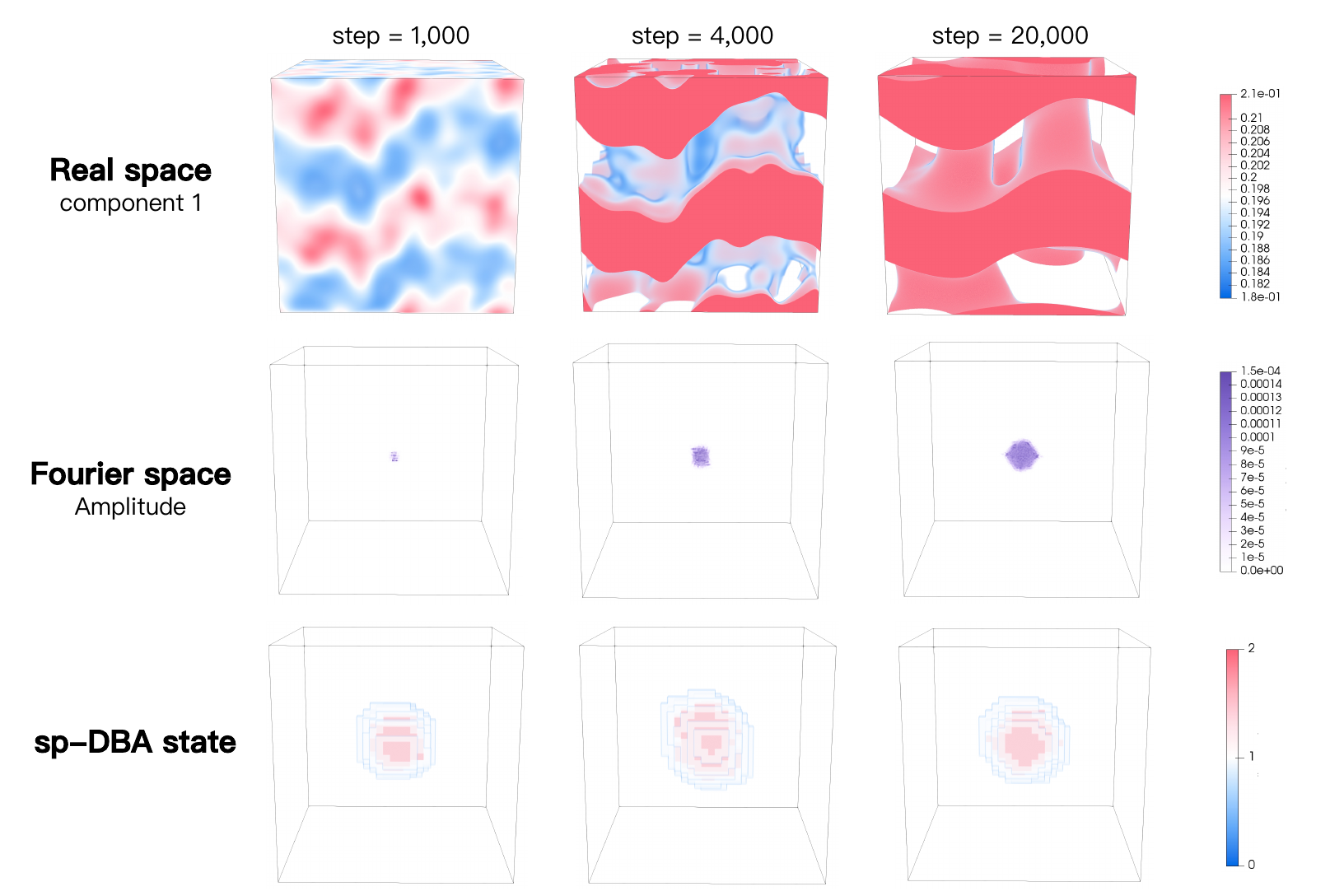}
\caption{Layered RCD pattern formation under sp-DBA on a \(256\times256\times256\) grid with \(M=8\). The top, middle, and bottom rows show one real-space component, Fourier amplitude, and sp-DBA block state, respectively. The average active-block ratio is about 3\%.}
\label{fig:s10_rcd_three_layer}
\end{figure}

The Fourier amplitude is localized in a compact low-wavenumber region, and the sp-DBA active blocks follow this region (Fig.~\ref{fig:s10_rcd_three_layer}).

\subsection{Matrix eigendecomposition for GPU acceleration}\label{supp:s4.3}
\label{sec:si_rcd_eigen}

For the constant-coefficient tests, the diffusion matrix is diagonalized once before time stepping,
\begin{equation} \mathbf{D}=\mathbf{V}\mathbf{\Lambda}\mathbf{V}^{-1}. \end{equation}
Then
\begin{equation} \mathbf{I}+\Delta t\,k^2\mathbf{D} = \mathbf{V} \left( \mathbf{I}+\Delta t\,k^2\mathbf{\Lambda} \right) \mathbf{V}^{-1}, \end{equation}
and the update becomes
\begin{equation} \hat{\bm{u}}^{\,n+1}(\bm{k}) = \mathbf{V} \left( \mathbf{I}+\Delta t\,k^2\mathbf{\Lambda} \right)^{-1} \mathbf{V}^{-1} \left[ \hat{\bm{u}}^{\,n}(\bm{k}) + \Delta t\,\widehat{\mathbf{R}}(\bm{u}^{\,n})(\bm{k}) \right]. \end{equation}
This form replaces the dense matrix inverse with two constant basis transforms and one diagonal scaling that depends on \(k\).

\begin{figure}[H]
\centering
\includegraphics[width=1.0\linewidth]{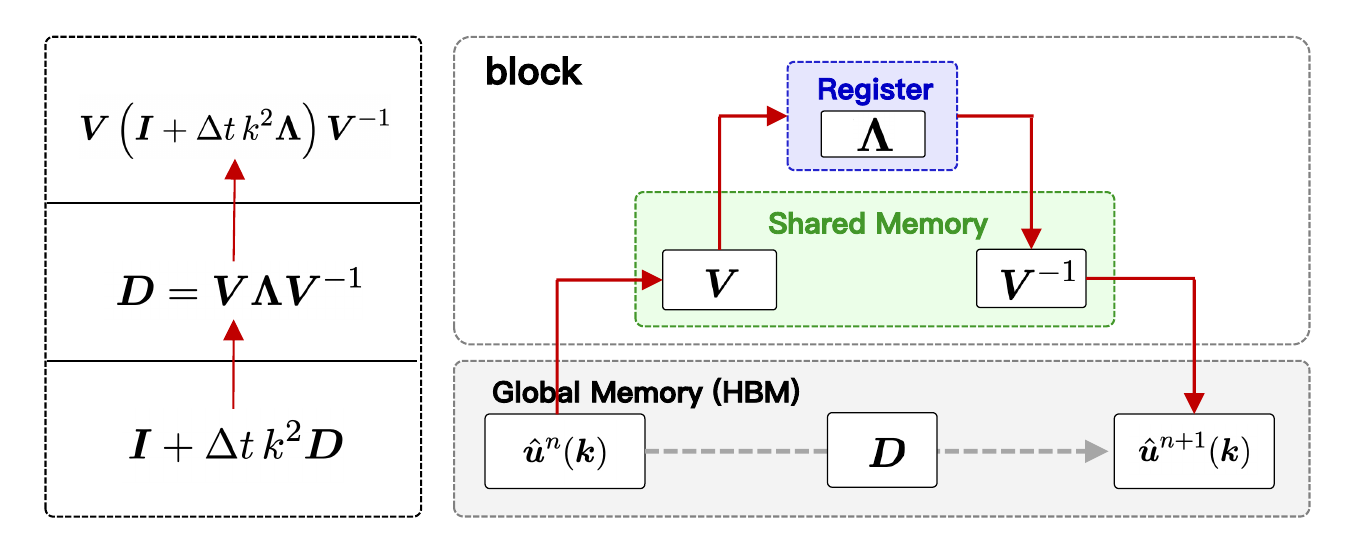}
\caption{Hardware-oriented diagonalized matrix update in the RCD solver. The gray path shows the direct matrix update, and the red path shows the reordered diagonalized implementation. The diagonal factor is held in registers, while the constant basis matrices are cached in shared memory for reuse within a block.}
\label{fig:s11_rcd_matrix}
\end{figure}

Eigendecomposition reduces the cost of the coupled update at each stored Fourier mode, whereas sp-DBA reduces the number of Fourier blocks where that update is executed. The two optimizations therefore act on different factors of the RCD update cost.

For the timing comparisons in the main text, Regular and sp-DBA use the direct coupled matrix update, whereas Regular + Eigen and sp-DBA + Eigen use the diagonalized update. The \(28.1\times\) value at \(M=48\) in Fig.~3d compares sp-DBA + Eigen with Regular using the direct matrix update; Fig.~5d reports all four configurations to separate the two contributions.

% ============================================================
\section{Digital backpropagation}
\label{sec:si_dbp}
% ============================================================

The DBP tests use a controlled split-step signal-recovery calculation motivated by space-division multiplexed optical communication~\cite{2008_DBP1,2009_DBP2}.

\subsection{Governing equations}\label{supp:s5.1}

Signal propagation is described by a simplified multimode model with linear mode coupling~\cite{2012_sdm_matrix,2013_manakov}. For the complex field envelope vector \(\bm{E}=[E_1,E_2,\ldots,E_M]^T\), the propagation equation is
\begin{equation} \frac{\partial \bm{E}}{\partial z}=(\hat{D}+\hat{N})\bm{E}. \end{equation}
The linear and nonlinear operators are
\begin{equation} \hat{D} = -\frac{\alpha}{2}\bm{I} -i\frac{\beta_2}{2}\frac{\partial^2}{\partial t^2}\bm{I} +i\bm{K}, \qquad \hat{N} = i\gamma \frac{8}{9} \left( \sum_{m=1}^{M}|E_m|^2 \right)\bm{I}. \end{equation}
Here, \(\alpha\) is the attenuation coefficient, \(\beta_2\) is the group-velocity dispersion coefficient, \(\gamma\) is the effective Kerr coefficient, and \(\bm{K}\) is a constant Hermitian mode-coupling matrix. The attenuation and dispersion terms act independently on each mode, while \(\bm{K}\) mixes the modal components through fixed linear coupling. The nonlinear operator gives the Manakov-averaged Kerr phase rotation driven by the total modal power and is evaluated pointwise in the time domain.

In the frequency domain, the dispersion part of \(\hat{D}\) is diagonal over frequency bins, while mode coupling acts as a constant matrix on the modal vector at each frequency. DBP approximates inverse propagation by reversing the dispersion, attenuation, and nonlinear parameters and by applying the inverse mode-coupling propagator~\cite{2008_DBP1,2009_DBP2}.

\subsection{Split-step Fourier implementation}\label{supp:s5.2}

Propagation is evaluated with a symmetric split-step Fourier method (SSFM)~\cite{1968_hfh_steps,2003_ssfm2,2024_sdm_nl_numericalAlg}. Over a step \(dz\),
\begin{equation} \bm{E}(z+dz,t)\approx \exp\!\left(\frac{dz}{2}\hat{D}\right) \exp\!\left(\int_z^{z+dz}\hat{N}(z')\,dz'\right) \exp\!\left(\frac{dz}{2}\hat{D}\right) \bm{E}(z,t). \end{equation}
Each propagation step contains two frequency-domain linear half-steps and one time-domain nonlinear step.

With both dispersion and mode coupling included in the linear operator, each linear half-step is factorized as
\begin{equation} \exp\!\left(\frac{dz}{2}\hat{D}\right) \approx \exp\!\left(i\bm{K}\frac{dz}{4}\right) \exp\!\left[ \left( -\frac{\alpha}{2}\bm{I} +i\frac{\beta_2}{2}\omega^2\bm{I} \right) \frac{dz}{2} \right] \exp\!\left(i\bm{K}\frac{dz}{4}\right), \end{equation}
where \(\omega\) is the angular frequency of the Fourier bin. The quarter-step coupling propagator is
\begin{equation} \bm{U}_{\mathrm{c}}(\delta)=\exp(i\delta\bm{K}), \qquad \delta=\frac{dz}{4}. \end{equation}
Because \(\bm{K}\) is Hermitian and constant, it is diagonalized once as
\begin{equation} \bm{K}=\bm{V}\bm{\Lambda}\bm{V}^{\dagger}, \qquad \bm{U}_{\mathrm{c}}(\delta)=\bm{V}\exp(i\delta\bm{\Lambda})\bm{V}^{\dagger}. \end{equation}
Forward and backward coupling actions are obtained by using opposite signs of \(\delta\), giving consistent discrete coupling propagators.

For the coupled-mode tests, $\bm{K}$ is generated once using \texttt{std::mt19937} with seed 1234.
Its diagonal entries are real samples from $\mathcal{N}(0,1)$; the real and imaginary parts of each upper-triangular off-diagonal entry are sampled independently from the same distribution, and the lower triangle is set by Hermitian symmetry.
The matrix is normalized such that
\begin{equation}
\frac{1}{M^2}\sum_{i,j=1}^{M}|K_{ij}|^2=1.
\end{equation}

sp-DBA is applied only to these linear half-steps. The FFT and IFFT calls, the time-domain nonlinear step, and the split-step ordering are unchanged.

\subsection{Simulation configuration and recovery metrics}\label{supp:s5.3}

\noindent\textbf{Simulation parameters.}

All propagation and signal parameters are expressed in the simulation units used throughout the DBP calculations.
The FFT size is varied from \(N_t=8192\) to \(262144\), and the number of coupled components is varied from \(M=2\) to \(M=64\) in the scaling tests. The propagation configuration uses a fiber length of 500~km, a step size of \(dz=0.1\)~km, and an output interval of 200 steps. The propagation parameters are \(\beta_2=-21\), \(\alpha=0.1\) in the exponential attenuation convention used by the simulation, and \(\gamma=1.3\).

For the sp-DBA runs, the frequency block size is \(B=32\), the threshold is initialized at \(10^{-4}\), and dynamic threshold control is enabled. Unless otherwise stated, the default performance setting targets a 10\% average active-block ratio. Additional accuracy and sensitivity tests use target average active-block ratios of 5\%, 10\%, 15\%, and 20\%. The threshold control interval is 50 steps, and the dynamic adjustment factor is 0.05. Reported active-block ratios are averages over the recovery run.

Deterministic attenuation changes the overall frequency-domain amplitude scale, so the dynamic threshold is rescaled by the running signal level before activation.

\noindent\textbf{Initial signal configuration.}

The transmitted field is initialized as a sum of three Gaussian wave packets with different carrier frequency offsets,
\begin{equation}
E_m(0,t)=\sum_{q=1}^{3} A_0 \exp\!\left[-\frac{(t-t_q)^2}{2w^2}\right] \exp\!\left(i\Omega_q t\right).
\end{equation}
The same initial waveform is assigned to all modal components. The three packets use temporal offsets \(t_q=(-20,0,20)\) and carrier offsets
\begin{equation}
\Omega_q=2\pi\frac{N_t}{T}(-0.15,0,0.15),
\end{equation}
where \(T=1000\) is the temporal window. Each packet has amplitude \(A_0=\sqrt{P_0/3}\), where \(P_0 = 1000\) is the prescribed total peak-power parameter, and \(w=10\) is the prescribed pulse width. 

\noindent\textbf{Frequency-block activation.}

For block selection, sp-DBA uses the total modal spectral power
\begin{equation} P(\omega)=\sum_{m=1}^{M}|\tilde{E}_m(\omega)|^2 . \end{equation}
A frequency block is activated when its spectral power exceeds the current threshold after the attenuation rescaling described above, and the resulting state is applied to the full \(M\)-component frequency-domain vector in that block.

\noindent\textbf{Reconstruction metrics.}

Let \(\bm{E}_{\mathrm{tx}}(t)\) denote the transmitted field and \(\bm{E}_{p}(t)\) the recovered field for \(p\in\{\mathrm{reg},\mathrm{spDBA}\}\).

For each recovered field, the normalized mean-squared error is evaluated against the transmitted reference field,
\begin{equation} \mathrm{NMSE}_{p}(\mathrm{dB}) = 10\log_{10} \left( \frac{ \sum_t \left\| \bm{E}_{p}(t)-\bm{E}_{\mathrm{tx}}(t) \right\|_2^2 }{ \sum_t \left\| \bm{E}_{\mathrm{tx}}(t) \right\|_2^2 } \right). \end{equation}
Here, the sum is taken over the sampled time points and \(\|\cdot\|_2\) denotes the Euclidean norm over the modal components.

The field correlation coefficient is
\begin{equation} \mathrm{Corr}_{p} = \frac{ \left| \sum_t \bm{E}_{p}(t)^{\dagger}\bm{E}_{\mathrm{tx}}(t) \right| }{ \left( \sum_t \left\| \bm{E}_{p}(t) \right\|_2^2 \right)^{1/2} \left( \sum_t \left\| \bm{E}_{\mathrm{tx}}(t) \right\|_2^2 \right)^{1/2} }, \end{equation}
where \((\cdot)^{\dagger}\) denotes the Hermitian transpose.

\begin{figure}[H]
\centering
\includegraphics[width=1.0\linewidth]{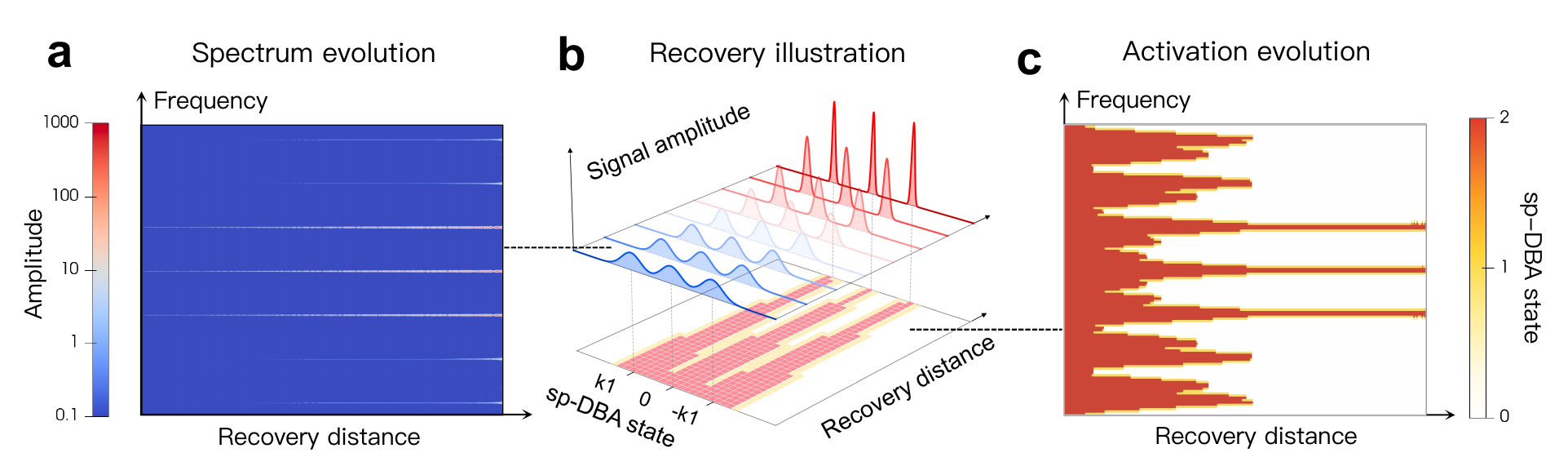}
\caption{Spectral evolution and sp-DBA activation in DBP. \textbf{a}, Spectrum versus recovery distance and frequency; color shows spectral amplitude. \textbf{b}, Schematic signal recovery and corresponding sp-DBA block states. \textbf{c}, Block-state evolution along the recovery distance. Low, intermediate, and high values correspond to frozen, transition, and active regions, respectively.}
\label{fig:s_dbp_spectrum_activation}
\end{figure}

The dominant spectral bands remain localized over much of the recovery distance, and the sp-DBA activation pattern follows their evolution (Fig.~\ref{fig:s_dbp_spectrum_activation}).

% ============================================================
% Keep S6 at section level in the global Contents.
\section{Numerical validation}
\label{sec:si_validation}
% ============================================================

\subsection{Conservation and free-energy evolution}\label{supp:s6.1}

For the XPFC application, we use the conserved mean-density mode and the free-energy trend to check whether block activation introduces systematic drift. The block containing the zero-frequency mode is kept active throughout the run, so the conserved mode is advanced as in Regular.

\begin{figure}[H]
\centering
\includegraphics[width=1.0\linewidth]{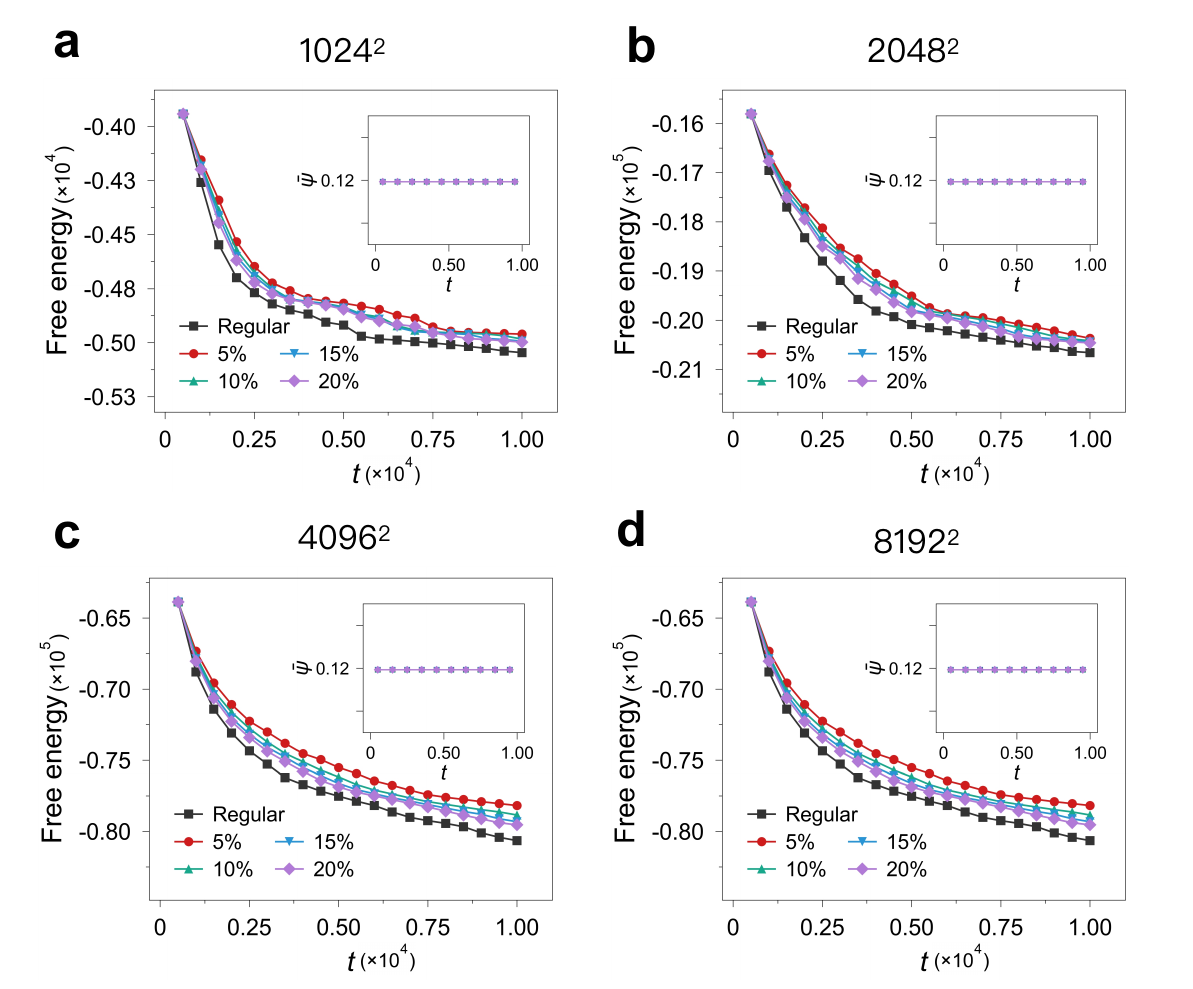}
\caption{Free-energy and mean-density agreement in the XPFC application under sp-DBA. \textbf{a--d}, Total free-energy evolution for Regular and sp-DBA on \(1024^2\), \(2048^2\), \(4096^2\), and \(8192^2\) grids. The insets show the conserved mean density \(\bar{\psi}\) for the corresponding runs.}
\label{fig:s9_conservation}
\end{figure}

Across the tested grid sizes, the mean density agrees with Regular, with no systematic increase in the free-energy discrepancy (Fig.~\ref{fig:s9_conservation}).

\subsection{Spectral accuracy and numerical error}\label{supp:s6.2}

\begin{figure}[H]
\centering
\includegraphics[width=1.0\linewidth]{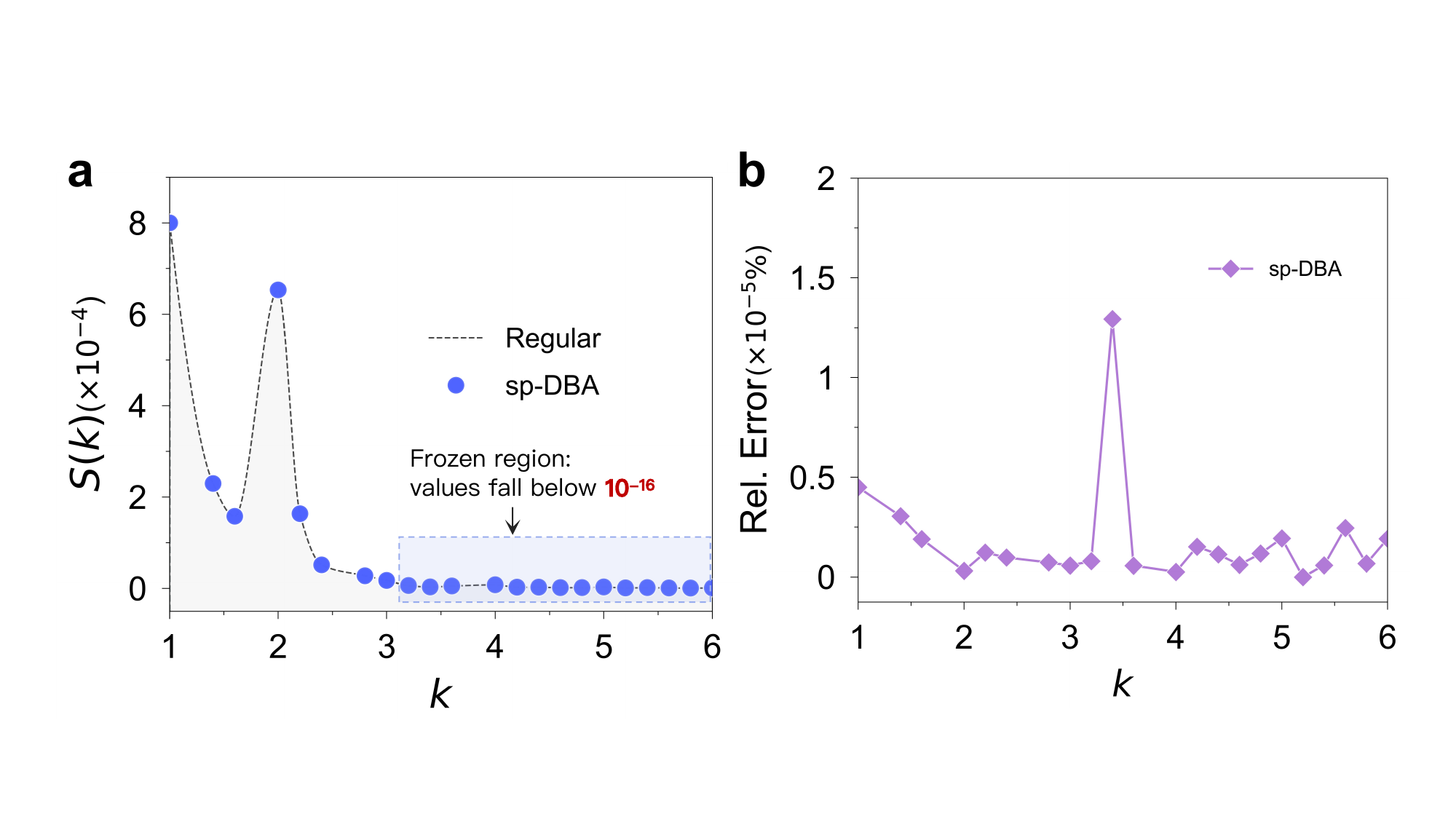}
\caption{Spectral agreement and numerical error under sp-DBA. \textbf{a}, Radial structure factor corresponding to the main-text comparison. Away from the principal peak region, the spectrum decays rapidly and the tail reaches the machine-precision level. \textbf{b}, Relative error after enabling sp-DBA.}
\label{fig:s10_spectral_error}
\end{figure}

Outside the principal peak region, the radial structure factor decays to approximately \(10^{-16}\), while the relative discrepancy after enabling sp-DBA remains about \(10^{-7}\) (Fig.~\ref{fig:s10_spectral_error}). This residual is consistent with the truncation-type error expected when spectrally small regions are retained from previous steps~\cite{2000_boyd}.

\subsection{Dealiasing compatibility}\label{supp:s6.3}

\begin{figure}[H]
\centering
\includegraphics[width=1.0\linewidth]{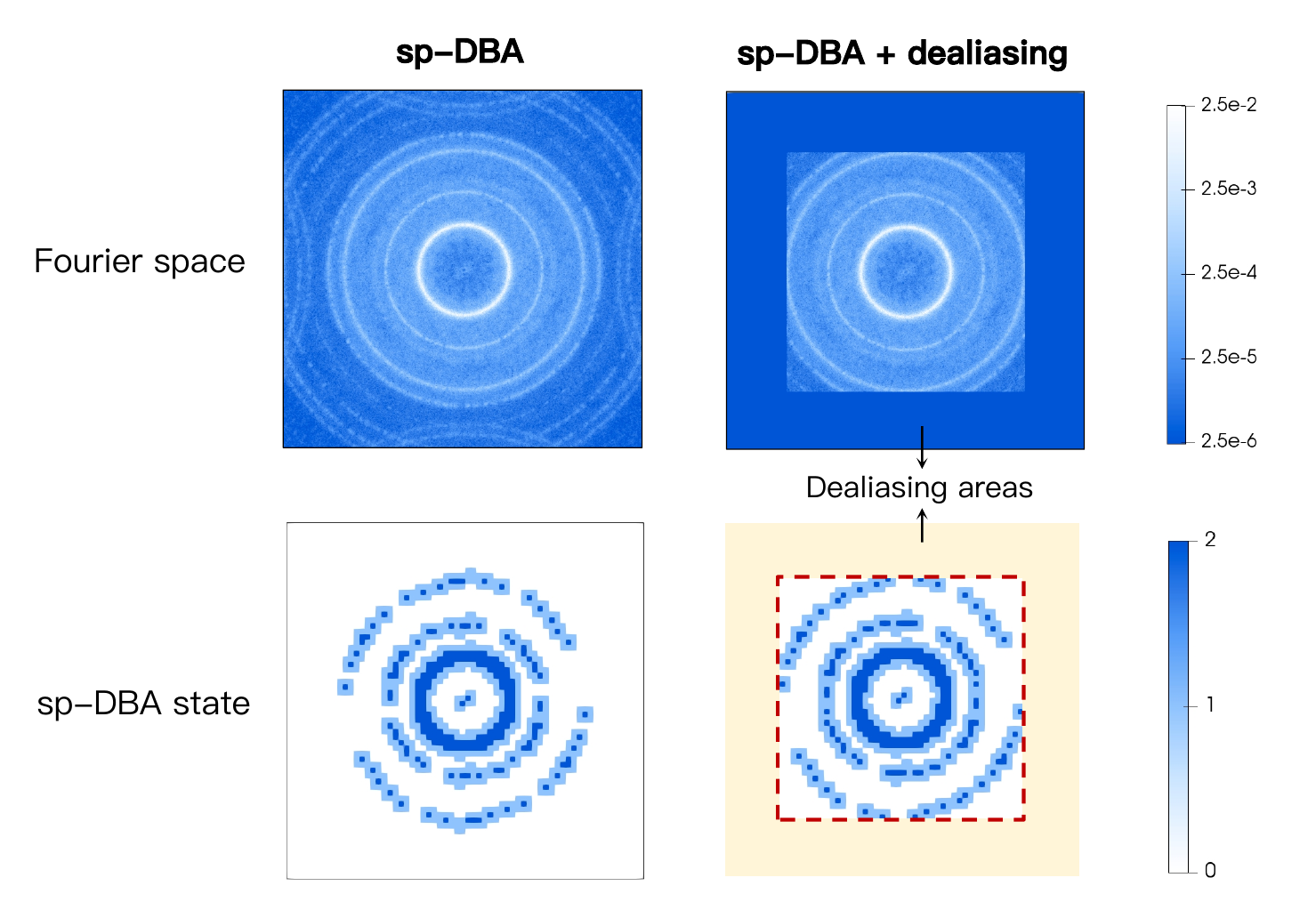}
\caption{Compatibility of sp-DBA with dealiasing. The two columns compare sp-DBA without and with dealiasing. In each column, the upper row shows the corresponding spectrum and the lower row shows the sp-DBA block activation state.}
\label{fig:s11_dealiasing}
\end{figure}

Pseudo-spectral evaluation of nonlinear terms can produce aliasing when unresolved high-frequency interactions fold back into the represented spectral band~\cite{2000_boyd}. Dealiasing first defines the retained spectral region, within which sp-DBA applies block activation (Fig.~\ref{fig:s11_dealiasing}).

% ============================================================
\addtocontents{toc}{\protect\setcounter{tocdepth}{2}}
\section{Single-GPU performance evaluation}
\label{sec:si_additional_performance}
% ============================================================

All single-GPU performance measurements were obtained on one NVIDIA A800-SXM4 GPU using the same CUDA
version, warm-up procedure, repeat count and GPU-synchronized wall-clock timing protocol listed for the distributed tests in Table~\ref{tab:dist_env}, with MPI disabled. Reported timings exclude file I/O and post-processing.

For runs in which communication is absent or not dominant, the measured end-to-end acceleration is approximated by
\begin{equation} S_{\mathrm{total}}= \frac{T_{\mathrm{reg}}}{T_{\mathrm{sp\mbox{-}DBA}}} \approx \frac{T_{\mathrm{reg}}} {T_{\mathrm{FFT}} + T_{\mathrm{NL}} + \alpha T_{\mathrm{update}} + T_{\mathrm{overhead}}}, \label{eq:si_total_speedup} \end{equation}
Here, \(T_{\mathrm{FFT}}\), \(T_{\mathrm{NL}}\), and \(T_{\mathrm{update}}\) denote the transform, retained model, and Regular update costs, respectively; \(\alpha\) is the average active-block ratio, and \(T_{\mathrm{overhead}}\) includes state evaluation, transition handling, refresh, and control operations. The unchanged workflow components impose the corresponding Amdahl-type bound~\cite{1967amdahl}.

A simpler estimate based on the active-block ratio is
\begin{equation} S_{\mathrm{estimate}}\approx \frac{T_{\mathrm{FFT}} + T_{\mathrm{NL}} + T_{\mathrm{update}}} {T_{\mathrm{FFT}} + T_{\mathrm{NL}} + \alpha T_{\mathrm{update}}}, \label{eq:si_theory_speedup} \end{equation}
where sp-DBA control overhead is omitted. Departures from this estimate arise from the omitted control overhead and hardware-dependent workload, memory, and execution effects. For the RCD comparison in Fig.~5e of the main text, Eq.~\eqref{eq:si_theory_speedup} serves as a block-activation estimate, while the measured curve also includes eigendecomposition.

\subsection{PFC cost decomposition and acceleration}\label{supp:s7.1}

\begin{figure}[H]
\centering
\includegraphics[width=1.0\linewidth]{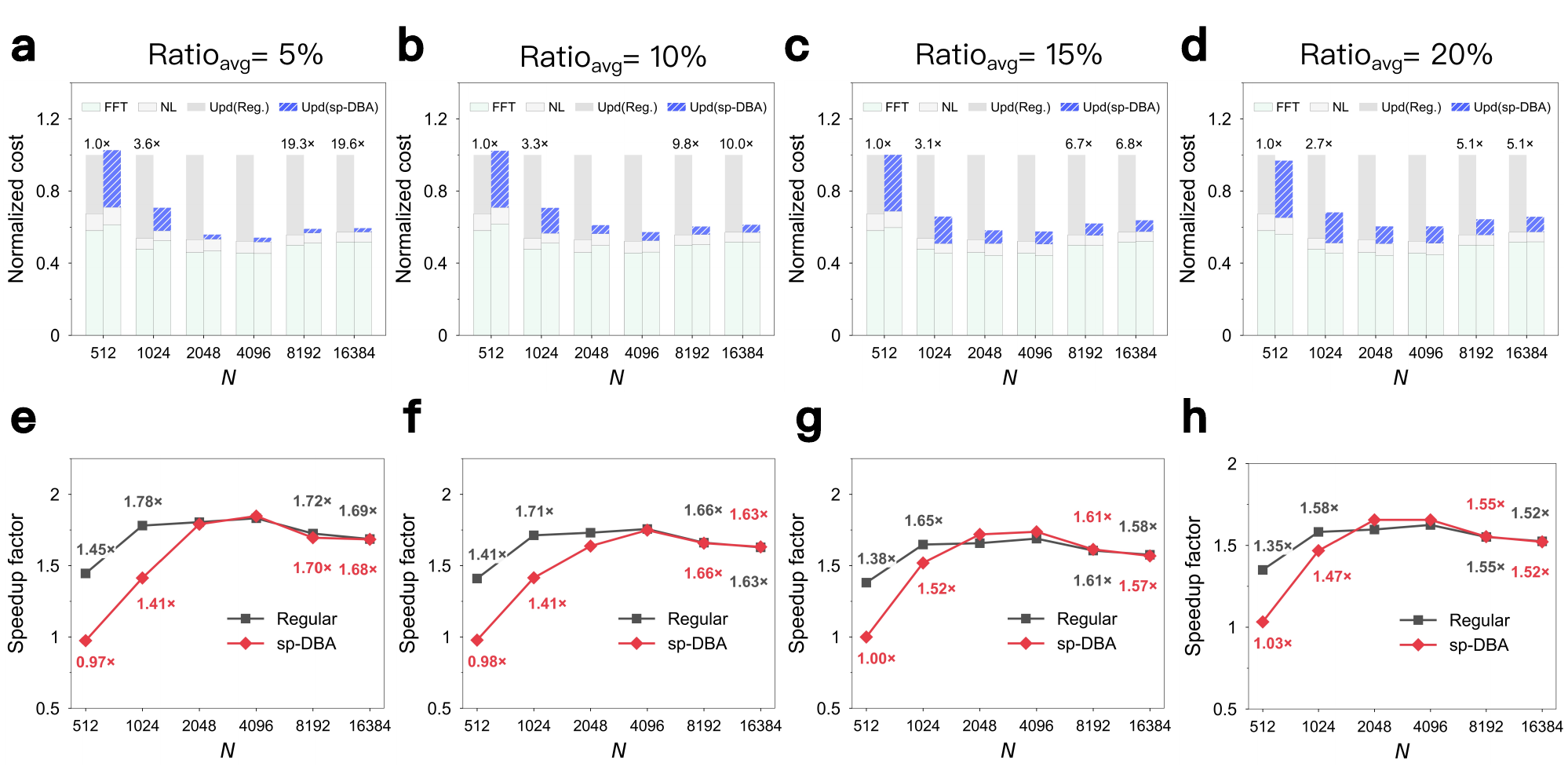}
\caption{Additional end-to-end performance analysis for the XPFC application. \textbf{a--d}, Total wall-clock cost decomposition for Regular and sp-DBA at average active-block ratios of 5\%, 10\%, 15\%, and 20\%, respectively. \textbf{e--h}, Comparison between the estimate from Eq.~\eqref{eq:si_theory_speedup} and the measured end-to-end acceleration for the same four active-block ratios.}
\label{fig:s12_pfc_performance}
\end{figure}

Most of the XPFC reduction comes from the transformed-field update, while the transform and retained model costs remain nearly unchanged (Fig.~\ref{fig:s12_pfc_performance}a--d). At larger grid sizes, the increased update workload better amortizes the block-control overhead, bringing the measured acceleration closer to the active-block-ratio estimate (Fig.~\ref{fig:s12_pfc_performance}e--h).

\subsection{DBP cost decomposition and acceleration}\label{supp:s7.2}

\begin{figure}[H]
\centering
\includegraphics[width=1.0\linewidth]{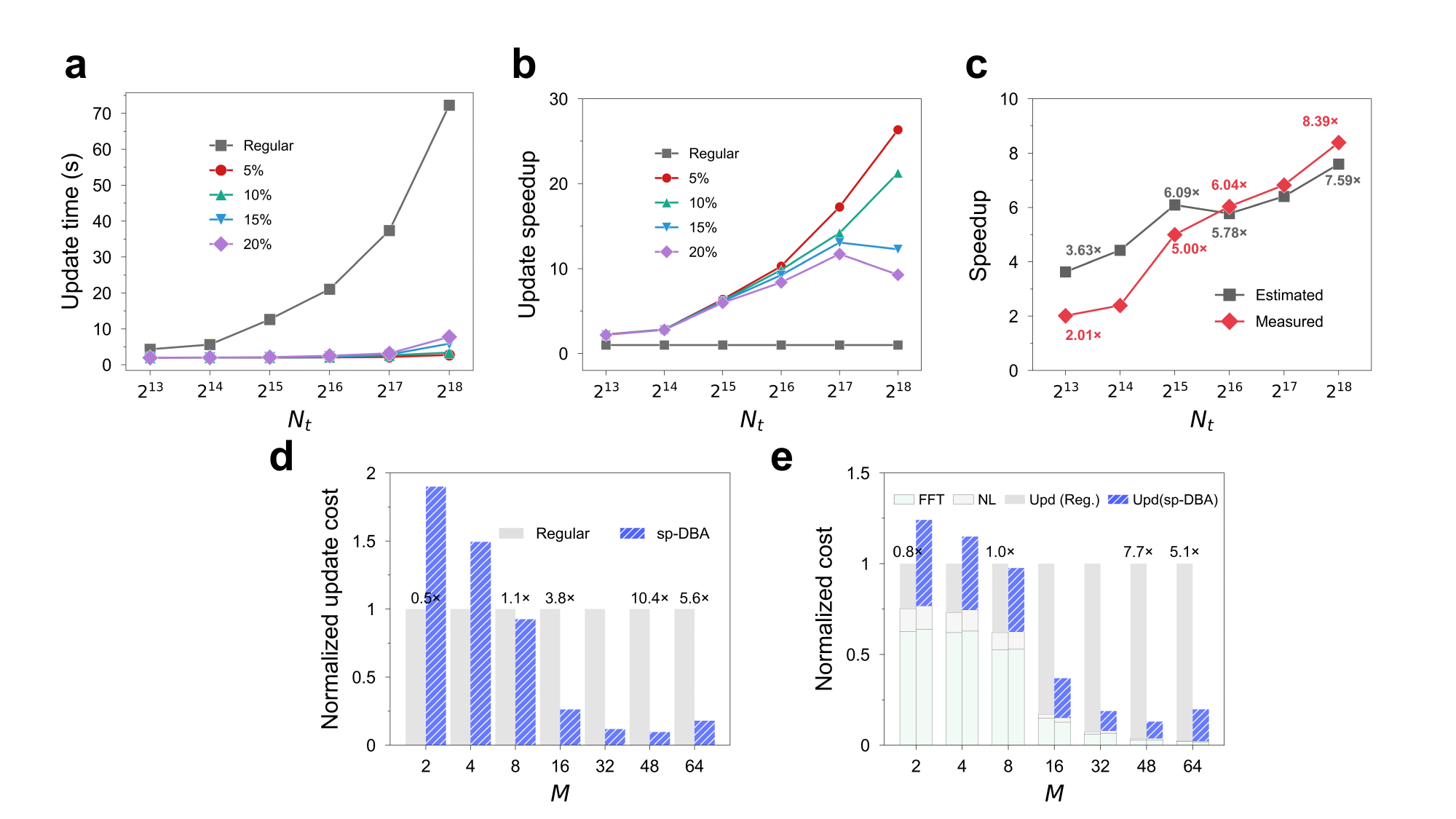}
\caption{Additional performance analysis for the DBP application. \textbf{a}, Frequency-domain update time as a function of FFT size \(N_t\) for Regular and sp-DBA runs with average active-block ratios of 5\%, 10\%, 15\%, and 20\%. \textbf{b}, Update-stage speedup for the same \(N_t\) sweep. \textbf{c}, Estimated and measured end-to-end acceleration as a function of \(N_t\). \textbf{d}, Normalized frequency-domain update cost as a function of the number of coupled components \(M\) for Regular and sp-DBA. \textbf{e}, Normalized total wall-clock cost decomposition as a function of \(M\), separating FFT, nonlinear time-domain step, Regular frequency-domain update, and sp-DBA frequency-domain update costs.}
\label{fig:s13_dbp_performance}
\end{figure}

As \(N_t\) increases, the Regular frequency-domain update grows faster than the sp-DBA update, giving larger update-stage speedups for lower active-block-ratio settings (Fig.~\ref{fig:s13_dbp_performance}a,b). Under the default 10\% setting used for the main-text performance summary, the update-stage speedup reaches \(21.3\times\) and the measured end-to-end acceleration reaches \(8.4\times\) at \(N_t=2^{18}\). The end-to-end acceleration is bounded by unchanged FFT/IFFT calls, the nonlinear step, and sp-DBA control overhead (Fig.~\ref{fig:s13_dbp_performance}c). The \(M\) sweep shows that the savings remain concentrated in the frequency-domain linear update (Fig.~\ref{fig:s13_dbp_performance}d,e).

% ============================================================
\section{Multi-GPU sp-DBA with distributed FFTs}
\label{sec:si_distributed}
% ============================================================

The distributed PFC tests apply sp-DBA to the local transformed-field update within a slab-decomposed FFT workflow. The distributed solver uses a semi-implicit pseudo-spectral update. These experiments assess distributed-workflow compatibility and update-stage scaling.
Unless stated otherwise, the physical parameters, block size, activation settings, and initialization follow Supplementary Section~S3.2.

\begin{table}[H]
\caption{Core settings for the distributed PFC tests.}
\label{tab:dist_env}
\centering
\begin{tabular}{@{}p{0.24\linewidth}p{0.70\linewidth}@{}}
\toprule
Item & Setting \\
\midrule
GPU platform & NVIDIA A800-SXM4 GPUs \\
CUDA / MPI & CUDA 12.2; Open MPI 4.1.6 with CUDA-aware device-buffer communication \\
Distributed FFT & Custom slab-decomposed 2D FFT implementation \\
Tested GPU counts & \(P=1,2,4,8\) \\
Strong-scaling grid & \(N=4096\) \\
Weak-scaling grids & \(N=2048,2880,4096,5760\) \\
Time steps & 10,000 \\
Repeats & Three repeated runs after warm-up \\
Timing & GPU-synchronized wall-clock time \\
\bottomrule
\end{tabular}
\end{table}

\subsection{Distributed FFT workflow}\label{supp:s8.1}

The distributed PFC implementation combines local cuFFT real-to-complex and complex-to-real transforms with MPI slab redistribution~\cite{cufft,2012_P3Dfft,2020_heffte}. Half-spectrum interpretation uses Hermitian symmetry and the remapping rule in Supplementary Section~\ref{sec:core_mechanisms}.

\begin{figure}[H]
\centering
\includegraphics[width=1.0\linewidth]{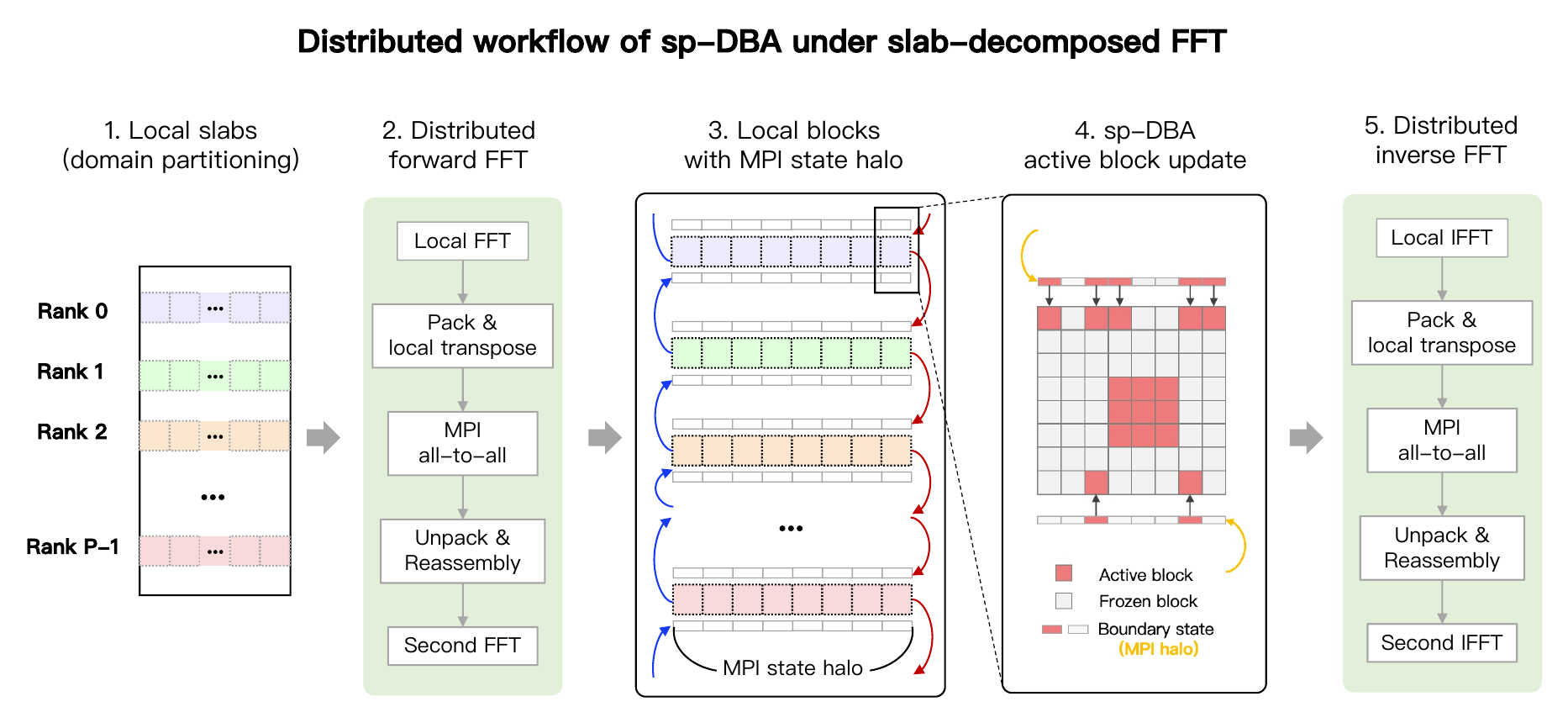}
\caption{Distributed workflow of sp-DBA under slab decomposition. The workflow consists of local slab partitioning, distributed forward FFT, local block construction with MPI state halos, sp-DBA transformed-field update, and distributed inverse FFT. The sp-DBA update is applied after each rank owns its assembled spectral data, while the forward and inverse distributed FFT stages retain the communication pattern required by the transform.}
\label{fig:s18_dist_workflow}
\end{figure}

\begin{figure}[H]
\centering
\includegraphics[width=1.0\linewidth]{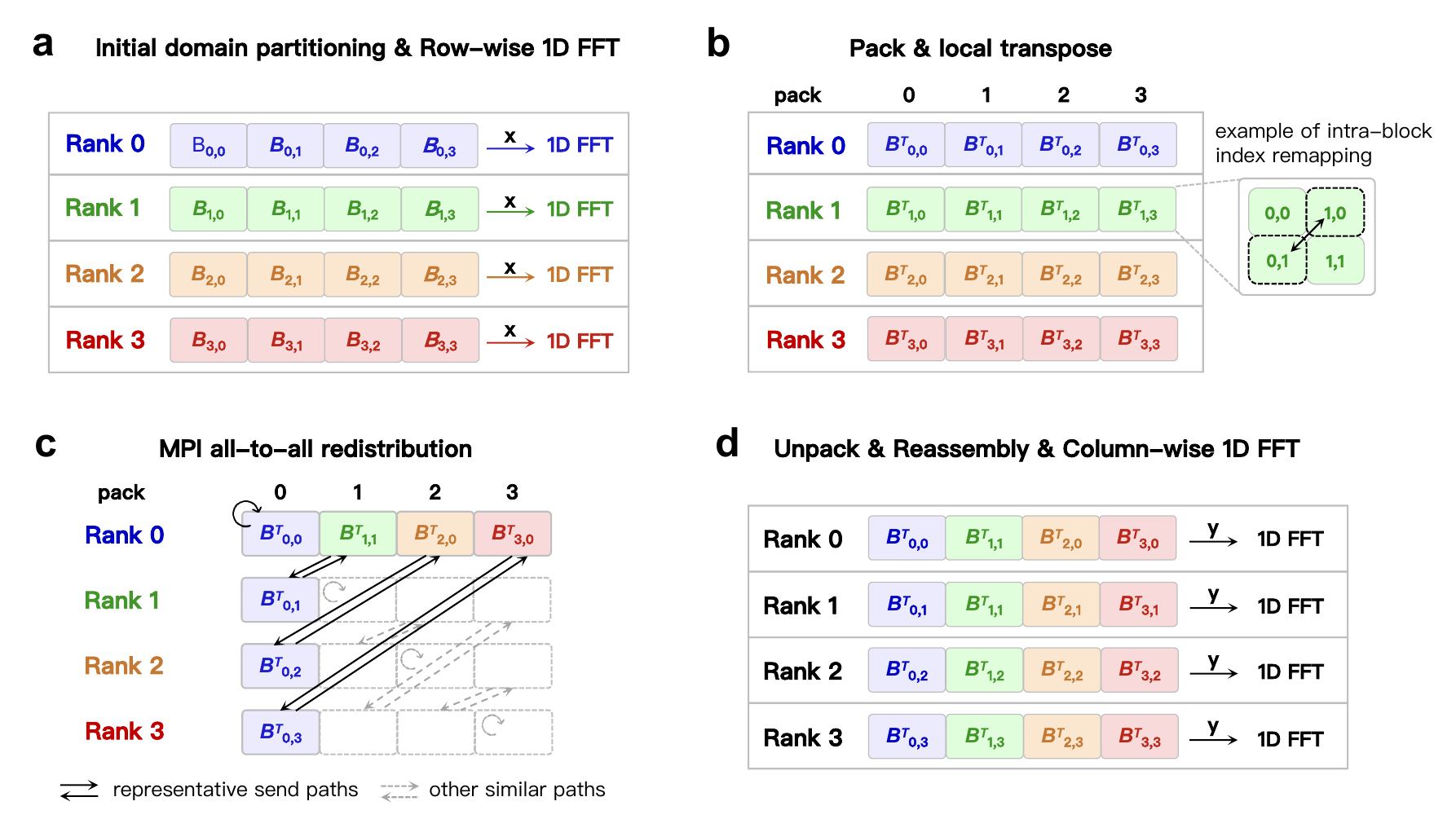}
\caption{Slab decomposition and block redistribution in the distributed two-dimensional FFT. \textbf{a}, Row-wise slab decomposition before the first local FFT. \textbf{b}, Local block organization and transpose before MPI redistribution. \textbf{c}, MPI all-to-all redistribution of transposed blocks. \textbf{d}, Reassembled layout used for the second one-dimensional FFT.}
\label{fig:s19_slab_decomposition}
\end{figure}

After the distributed FFT assembles the spectral data on each rank, each rank performs the local sp-DBA update and exchanges only the block states required at rank boundaries (Fig.~\ref{fig:s19_slab_decomposition}); the global transpose and all-to-all redistribution remain unchanged.

\begin{figure}[H]
\centering
\includegraphics[width=1.0\linewidth]{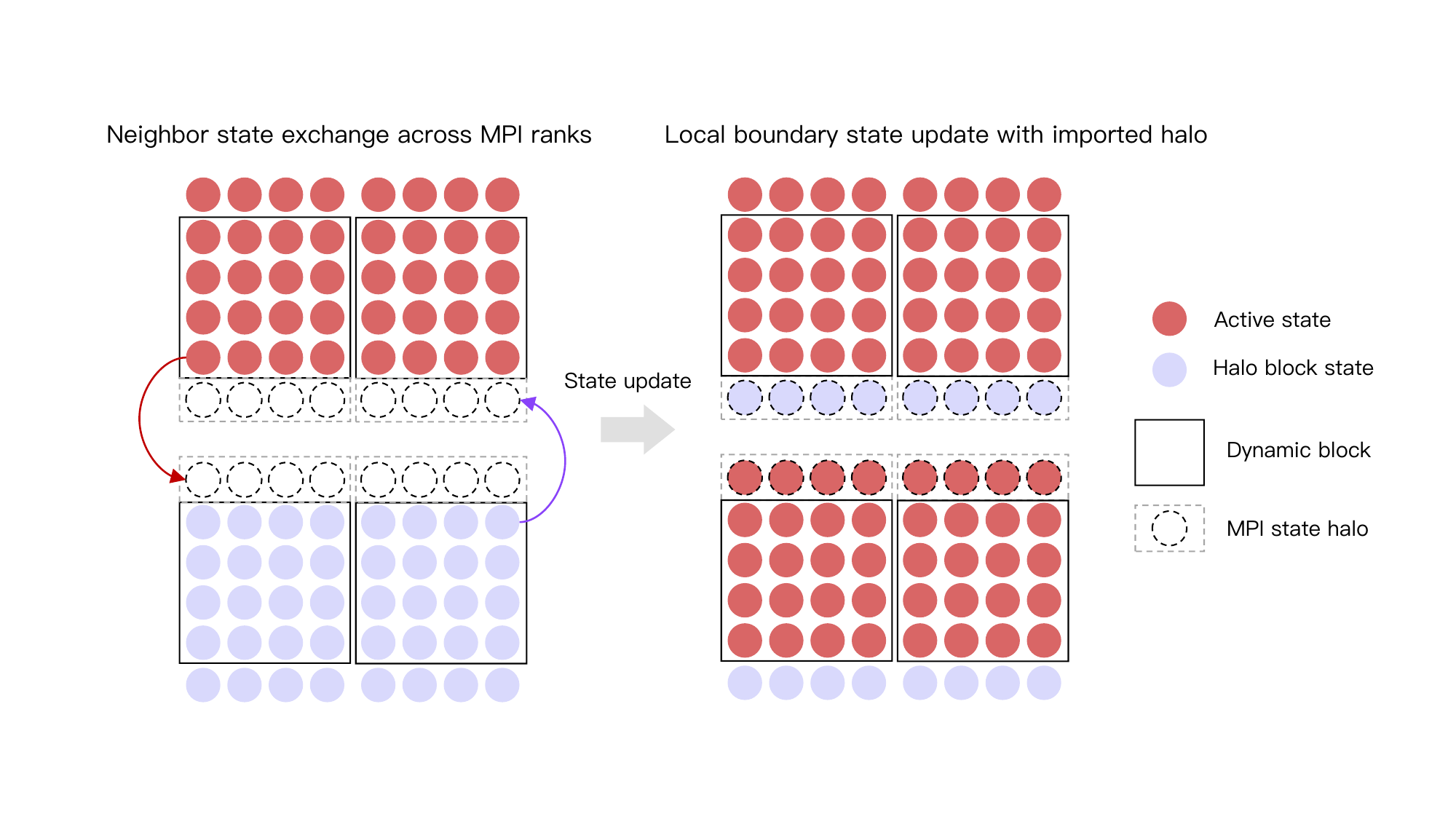}
\caption{Rank-boundary metadata exchange for distributed sp-DBA. Block states at neighboring rank boundaries are exchanged as metadata. The imported halo states are then used to update local boundary states so that active and transition blocks are assigned consistently across rank partitions.}
\label{fig:s20_metadata_exchange}
\end{figure}

 The distributed implementation uses tiled slab packing, self-rank bypass, CUDA-aware nonblocking MPI, and device-buffer metadata exchange, while the sp-DBA update mechanism itself is unchanged.

\subsection{Strong and weak scaling}\label{supp:s8.2}

\begin{figure}[H]
\centering
\includegraphics[width=1.0\linewidth]{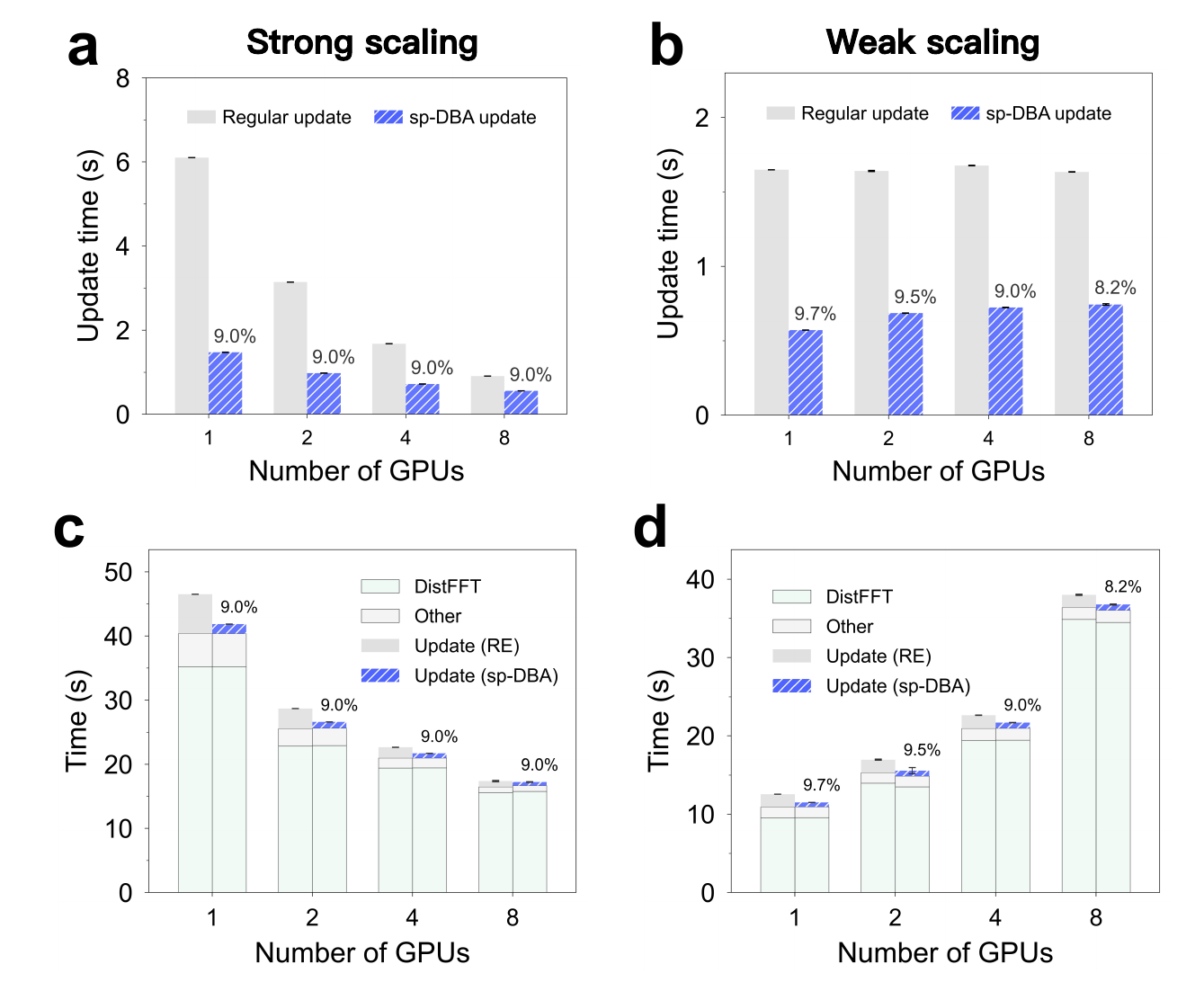}
\caption{Distributed scaling of sp-DBA in the PFC application. \textbf{a}, Update stage under strong scaling at fixed global grid size \(N=4096\). \textbf{b}, Update stage under weak scaling with approximately fixed local grid area per GPU. \textbf{c}, Overall strong-scaling wall-clock decomposition at fixed global grid size \(N=4096\). \textbf{d}, Overall weak-scaling wall-clock decomposition with approximately fixed local grid area per GPU. Gray and blue bars denote the Regular and sp-DBA update stages, respectively; numbers above the blue bars indicate the measured average active-block ratio, not speedup.}
\label{fig:s21_dist_scaling}
\end{figure}

Under strong scaling, both update times decrease with the local workload, with the sp-DBA update time below the Regular update time (Fig.~\ref{fig:s21_dist_scaling}a). Under weak scaling, the sp-DBA update time changes little and the measured average active-block ratio stays close to 10\% (Fig.~\ref{fig:s21_dist_scaling}b).

Distributed FFT and communication dominate the total wall-clock cost (Fig.~\ref{fig:s21_dist_scaling}c,d), so the overall acceleration is bounded by the share of the local update in the distributed workflow.

\subsection{CUDA-aware MPI communication}\label{supp:s8.3}

\begin{figure}[H]
\centering
\includegraphics[width=1.0\linewidth]{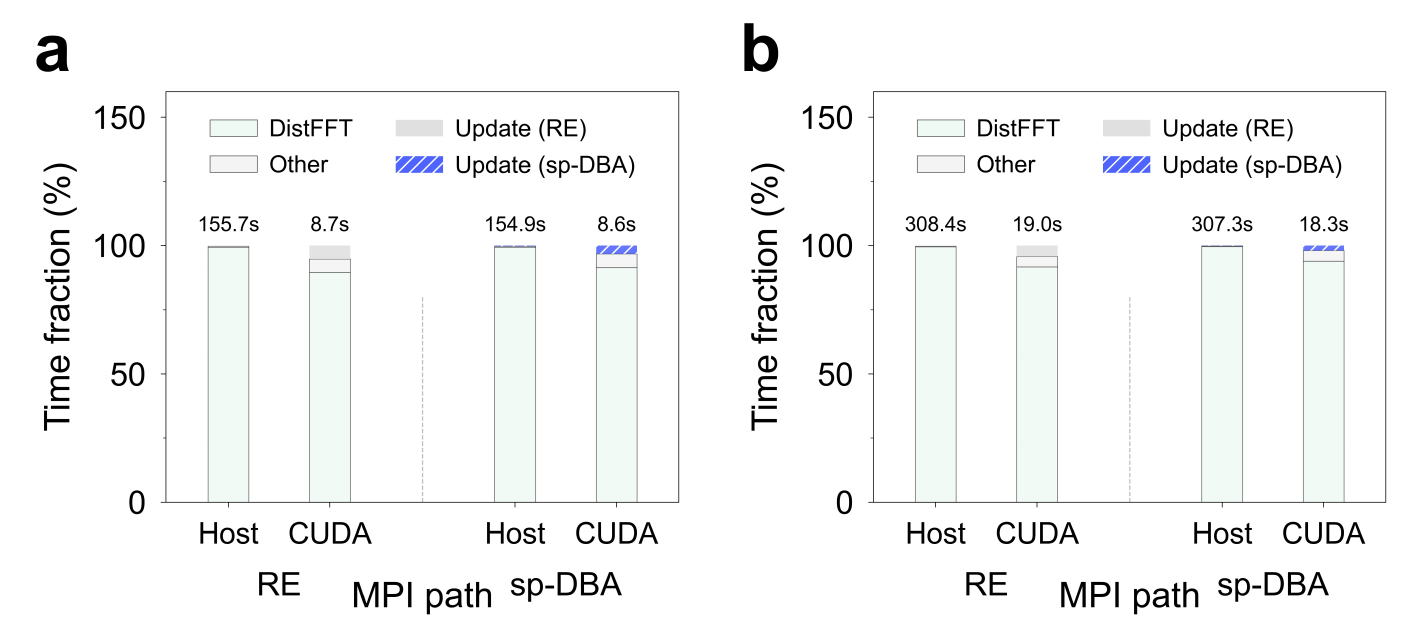}
\caption{Communication path comparison between host staging and CUDA-aware MPI. \textbf{a}, Strong-scaling control case at \(P=8\) and \(N=4096\). \textbf{b}, Weak-scaling control case at \(P=8\) and \(N=5760\). Within each solver mode, the host-staged path is shown on the left and CUDA-aware device-buffer MPI is shown on the right. Numbers above the bars indicate total wall-clock time.}
\label{fig:s22_cuda_aware}
\end{figure}

Reported timings use CUDA-aware device-buffer MPI; host-staged runs serve only as communication-path controls (Fig.~\ref{fig:s22_cuda_aware}).

\subsection{Activation stability under distributed execution}\label{supp:s8.4}

\begin{figure}[H]
\centering
\includegraphics[width=1.0\linewidth]{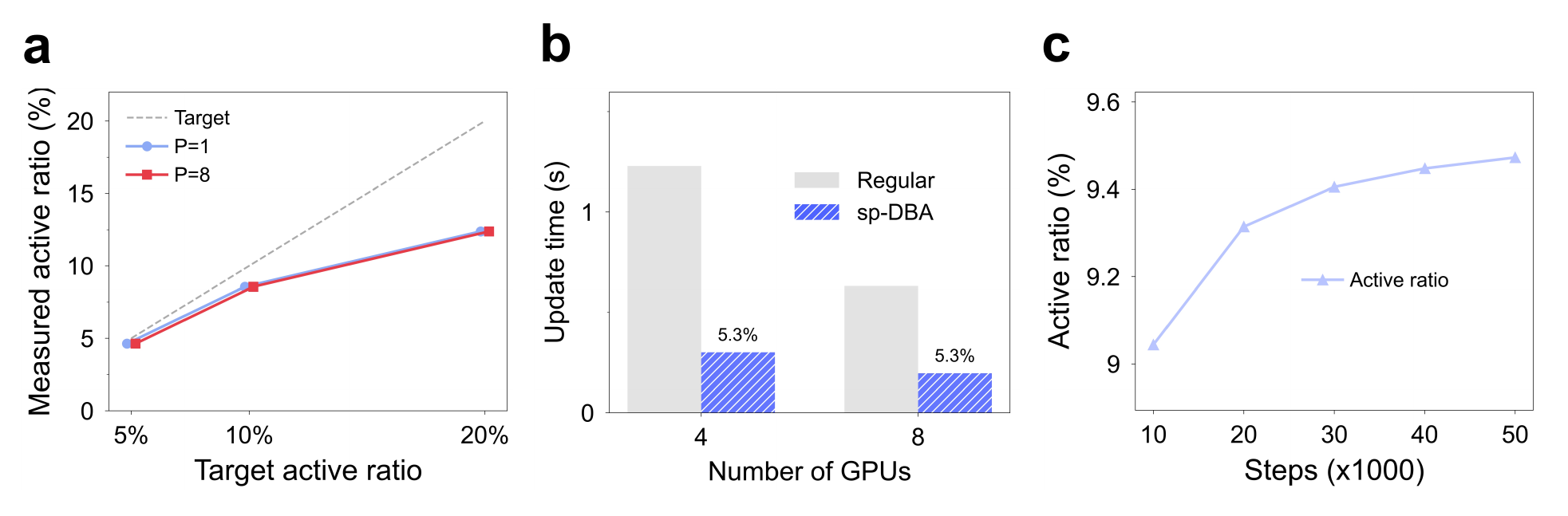}
\caption{Activation stability in distributed PFC tests. \textbf{a}, Measured active-block ratio as a function of the target average active-block ratio at \(P=1\) and \(P=8\). The dashed line denotes exact target tracking. \textbf{b}, Large-\(N\) update-stage comparison at \(N=8192\) on 4 and 8 GPUs. Gray and blue bars denote the Regular and sp-DBA update times, respectively; numbers above the blue bars indicate the measured average active-block ratio. \textbf{c}, Average active-block ratio during a 50,000-step distributed sp-DBA run at \(P=8\) and \(N=4096\).}
\label{fig:s23_activation_stability}
\end{figure}

The target-ratio test varies the target average active-block ratio from 5\% to 20\%. The measured ratio follows the target but remains below the exact tracking line in the distributed setting (Fig.~\ref{fig:s23_activation_stability}a). The 10\% setting is used as the default because it provides a clear update-time reduction while keeping the measured ratio near the target range.

At \(N=8192\), sp-DBA reduces the local transformed-field update time on both 4 and 8 GPUs relative to Regular (Fig.~\ref{fig:s23_activation_stability}b).

Over 50,000 steps at \(P=8\) and \(N=4096\), the active-block ratio stays close to the default 10\% range, indicating stable dynamic threshold control and rank-boundary state exchange under the tested decomposition (Fig.~\ref{fig:s23_activation_stability}c).

% ============================================================
\addtocontents{toc}{\protect\setcounter{tocdepth}{1}}
\section{Visualization}
\label{sec:si_visualization}
% ============================================================

ParaView~\cite{paraview} was used for field, Fourier-amplitude, isosurface, and volume rendering, while OVITO~\cite{ovito} was used for XPFC grain reconstruction and morphology visualization. Activation maps were generated from recorded block labels, while scalar quantities and performance plots were prepared from exported analysis files. Post-processing was excluded from solver timing unless stated otherwise.

\end{document}